\documentclass[preprint]{revtex4-1}
\usepackage{fontenc}[T1]
\usepackage{amsmath}
\usepackage{amssymb}
\usepackage{graphicx}
\usepackage{color}
\usepackage[usenames]{xcolor}

\usepackage{helvet}
\usepackage{txfonts}
\usepackage{isomath}
\usepackage{upgreek}
\usepackage{bm}
\usepackage{enumerate}
\usepackage{braket}

\newcommand{\ci}{i}
\newcommand{\cpi}{\uppi}
\newcommand{\ce}{e}
\newcommand{\dd}{d}

\newcommand{\average}[1]{\langle#1\rangle}
\renewcommand{\Im}{\operatorname{Im}}
\renewcommand{\Re}{\operatorname{Re}}
\newcommand{\sgn}{\operatorname{sgn}}

\newcommand{\abs}[1]{\lvert\,#1\,\rvert}
\newcommand{\cconj}[1]{{#1}^{*}}
\newcommand{\hconj}[1]{{#1}^{\dag}}
\newcommand{\commute}[2]{\left[#1,\:#2\right]}

\newcommand{\Order}[1]{\mathrm{O}\left({#1}\right)}

\newcommand{\LeftClosedRightOpenInterval}[2]{[#1,#2)}

\newcommand{\citen}[1]{\cite{#1}}

\begin{document}
\title{Effects of Weak and Strong Scatterers on the spectra of Vortex Andreev Bound States in Two-Dimensional Chiral \textit{p}-wave Superconductors}
\author{Noriyuki Kurosawa}
\affiliation{Department of Basic Science, The University of Tokyo, Komaba, Meguro, Tokyo 153-8902, Japan}
\author{Nobuhiko Hayashi}
\affiliation{NanoSquare Research Center (N2RC), Osaka Prefecture University, 1-2 Gakuen-cho, Naka-ku, Sakai 599-8570, Japan}
\author{Yusuke Kato}
\affiliation{Department of Basic Science, The University of Tokyo, Komaba, Meguro, Tokyo 153-8902, Japan}
\date{\today}

\begin{abstract}
  The vortices of two-dimensional chiral \textit{p}-wave superconductors are predicted to exhibit some exotic behaviors; one of their curious features is the existence of two types of vortices (each vortex has vorticity either parallel or antiparallel to the Cooper pair's chirality) and the robustness of the antiparallel vortices against nonmagnetic Born-like impurities.
  In this work, we study the impurity effect on the vortex of the chiral \textit{p}-wave superconductors through the quasiclassical Green's function formalism. We take account of impurities via the self-consistent \textit{t}-matrix approximation so that we can deal with strong as well as Born-like (i.e., weak) scatterers.
  We found that the spectrum is heavily broadened when the phase shift $\delta_0$ of each impurity exceeds a critical value $\delta_{\text{c}}$ above which the impurity band emerges at the Fermi level. We also found a quantitative difference in the impurity effects on the two types of vortex for $\delta_0<\delta_{\text{c}}$. Part of the numerical results for $\delta_0<\delta_{\text{c}}$ can be understood by a variant of the analytical theory of Kramer and Pesch for bound states localized within vortex cores.
\end{abstract}

\maketitle

\section{Introduction}

Low-energy excitations localized within quantized vortices of second-kind superconductors, known as the Caroli-de~Gennes-Matricon (CdGM) mode\cite{Caroli1964}, dominate the physics of vortices and the mixed phase of superconductors; these states considerably affect vortex flow resistivity, critical current, and  thermodynamic quantities at temperatures much lower than the transition temperature.
 In the usual case, the CdGM mode is well understood as a kind of the Andreev bound states (ABS)\cite{Stone1996}, and we refer to the bound-states as vortex-ABS in this paper.

Spin triplet superconductors (SCs) or superfluids (SFs), such as superfluid helium-3, exhibit various phenomena that stem from their spin and orbital degrees of freedom. The vortices of these superconductors can show a variety of behaviors that never take place in conventional $s$-wave superconductors. Their rich physics has attracted not only theoretical but also practical interest.
For example, it has been predicted that a zero-energy state that obeys non-Abelian statistics can emerge within the vortices of some types of spin-triplet superconductors\cite{Read2000, Ivanov2001}. These exotic systems are considered to be a candidate platform of novel devices for quantum computing\cite{Nayak2008}.

A chiral $p$-wave SC/SF is believed to be realized in the thin film of the helium-3 A-phase\cite{Vorontsov2003,Levitin2013} or Sr$_2$RuO$_4$\cite{Maeno1994, Mackenzie2003, Sigrist2005, Maeno2012}. This SC/SF has degenerated ground states that break time-reversal symmetry, and accordingly, the Cooper pair of this system has a nonzero internal angular momentum. This unusual feature causes unique phenomena such as the internal Magnus effect\cite{Salmelin1989, Salmelin1990, Ikegami2013}.

Another anomalous feature caused by the internal angular momentum is the existence of two different types of vortex; each vortex has angular momentum (vorticity) either parallel or antiparallel to the internal angular momentum (chirality) of the Cooper pairs. In this paper, we call the former  a ``parallel vortex'' and the latter an ``antiparallel vortex''\cite{Tanuma2009}.

There have been some theoretical predictions that parallel and antiparallel vortices exhibit qualitative differences\cite{Tanuma2009, Tokuyasu1990, Matsumoto2000, Matsumoto2001,Kato2000, Kato2002, Hayashi2005}. One of the most significant differences is the impurity effect on the vortex-ABS; the antiparallel vortex is very tolerant toward impurities but the parallel one is not\cite{Matsumoto2000, Kato2000, Kato2002, Hayashi2005, Tanuma2009}.
This robustness is unique to the antiparallel vortex of the chiral $p$-wave superconductors and is very useful in the sense that it will enable us to detect the domains of the degenerate states using vortices. The connection between this phenomenon and the topologically protected states or the odd-frequency pairing\cite{Tanuma2009, Yokoyama2009, Tanaka2012, Volovik1999} has been widely discussed.

Most of the previous studies, however, have been carried out on the effect of a single impurity within a vortex core\cite{Volovik1999, Matsumoto2000} or spatially averaged Born-like (weak) impurities \cite{Kato2000, Kato2002, Hayashi2005, Tanuma2009}. In some systems, impurities are very strong and should be treated not as Born-like but as unitary-like (e.g.,\ Refs.\ \citen{Hirschfeld1986, SchmittRink1986, Suzuki2002}), and hence previous studies are insufficient from this viewpoint.
Indeed, the Kramer-Pesch effect\cite{Kramer1974} of antiparallel vortices is weakened by sufficiently strong scatterers\cite{Hayashi2013-pwave}, and other properties of a vortex may also be different between Born- and unitary-like scatterers.
Generally speaking, unlike weak scatterers, strong scatterers within anisotropic superconductors generate impurity bands within the superconducting gap. Chiral $p$-wave superconductivity is fully gapped, and thus the effect of the emergence of low-energy excitation states by the impurities on this system is expected to be more apparent than that in nodal superconductors.

In addition, even for the Born-like impurities, Sauls and Eschrig have reported that the difference between parallel and antiparallel vortices is not so large under a self-consistent numerical calculation\cite{Sauls2009}.
A consensus regarding the impurity effect on this system has not yet been established.

Thus, we study the impurity effect on the vortex of the chiral $p$-wave superconductors, which are expected to have many anomalous features, some of which are still controversial.
In a previous work\cite{Kurosawa2014} we reported the relationship between the phase-shift of the single impurity scattering and the local density of states (LDOS) within an isolated integer vortex through numerical calculations. In this paper, we will show more numerical results and interpretations. 

This paper is organized as follows. In Sec.~2, we provide a brief summary of the  quasiclassical method used in this study. In Sec.~3, we present the results of our numerical calculations. In Sec.~4, we discuss the physics and try to explain it analytically with the Kramer-Pesch approximation, and the last section is for concluding remarks.
Throughout this paper, we use a unit system where $\hbar = 1$ and $k_{\text{B}}=1$. 

%
\section{Formulation}
\label{section:formulation}
\subsection{Quasiclassical theory of superconductivity}
\label{subsection:quasiclassical}
We consider two-dimensional chiral $p$-wave superconducting systems with a circular symmetric Fermi surface. The London penetration depth of the superconductor is assumed to be much larger than the coherence length, and thus we neglect the electromagnetic potential. 

In this study, we use the theory so-called quasiclassical method\cite{Eilenberger1968, Larkin1969, Serene1983}. This theory corresponds to the Andreev approximation and is well suited to study most superconducting systems, where the coherence length $\xi_0$ at zero temperature in the absence of impurities is much larger than its inverse Fermi momentum $k_{\text{F}}^{-1}$.
The quasiclassical Green's function $\check g$ is defined as
\begin{align}
  \check g(z, \bm{r}, \bm{\hat k}) =
  \int\dd\varepsilon_{\bm{k}}\check G(z, \bm{r},\bm{k}) =
  \begin{pmatrix}
    g & f \\ -\hconj{f} & -g
  \end{pmatrix}, 
  \label{eq:definition-quasiclassical-Green-function}
\end{align}
where $\check G$ is the spinless Gor'kov Green's function, $\bm{\hat k}$ is a unit vector in the direction of the Fermi momentum defined as $\bm{k}=k_{\text{F}}\bm{\hat k}=(k_{\text{F}}\cos\alpha, k_{\text{F}}\sin\alpha)$, $\varepsilon_{\bm{k}} = k^2/(2m^{*})$, where $m^{*}$ is the mass of an electron, and $z$ is a complex frequency that is taken as a Matsubara frequency $\ci\omega_n = \ci\cpi T(2n+1)$ when we search for the pair potential in the self-consistent calculation, or a real frequency $\epsilon+\ci\eta$ when we calculate the spectral function of the retarded Green's functions. The quasiclassical Green's function satisfies the normalized condition $\check g^2 = -\cpi^2\check \tau_0$ and obeys the Eilenberger equation\cite{Eilenberger1968}

\begin{align}
  -\ci\bm{v}_{\text{F}}\cdot\bm{\nabla}\check g = \commute{z\check \tau_3-\check\Delta-\check\Sigma}{\check g}
  .
  \label{eq:Eilenberger-equation}
\end{align}
Here, $\commute{\check A}{\check B} = \check A\check B-\check B\check A$ is the commutator, and $\bm{v}_{\text{F}}=v_{\text{F}}(\cos\alpha, \sin\alpha)$ denotes the Fermi velocity. $\check\tau_i$ ($i=0,1,2,3$) are the Pauli matrices in the particle-hole space. $\check\Delta$ is the pair potential
\begin{align}
  \check\Delta(\bm{r},\bm{\hat k}) = \begin{pmatrix}
    0 & \Delta(\bm{r},\bm{\hat k})\\
    -\cconj{\Delta}(\bm{r},\bm{\hat k}) & 0
  \end{pmatrix}
\end{align}
and
\begin{align}
  \check\Sigma(z, \bm{r}) = \begin{pmatrix}
    \Sigma_{\text{d}}(z, \bm{r}) & \Sigma_{12}(z, \bm{r})\\
    \Sigma_{21}(z, \bm{r}) & -\Sigma_{\text{d}}(z, \bm{r})
  \end{pmatrix}
\end{align}
is the impurity self-energy. We apply $t$-matrix approximation to obtain the self-energy from the Green's function\cite{Thuneberg1984},
\begin{align}
  \check\Sigma(z,\bm{r}) = \frac{\Gamma_{\text{n}}\average{\check g}/\cpi}{\cos^2\delta_0-\sin^2\delta_0\left(\average{g}^2-\average{f}\average{\hconj{f}}\right)/\cpi^2}
  \label{eq:self-energy-green-function}
  .
\end{align}
Here, $\Gamma_{\text{n}}$ is the scattering rate of the normal state and $\delta_0 \in [0,\cpi/2]$ is the phase-shift of a single impurity. The Born limit (the unitary limit) corresponds to $\delta_0=0$ ($\delta_0=\cpi/2$).
In this paper, we denote an averaged value on the Fermi surface by $\average{\bullet}$, i.e., $\average{A} = \average{A(\alpha)}_\alpha = \int\dd\alpha A(\alpha)/(2\cpi)$. We consider the partial wave with the angular momentum $\ell=0$ (\textit{s}-wave scattering) of impurity scattering and we ignore the anisotropic part (the partial wave with the angular momentum $\ell\ge 1$) .

The pair potential $\Delta$ satisfies the gap equation
\begin{align}
  \Delta(\bm{r},\bm{\hat k}) = N_0T\sum_{n, \abs{\omega_n}\le\omega_{\text{c}}}\average{U(\alpha-\alpha')f(\ci\omega_n,\bm{r},\alpha')}_{\alpha'}, 
  \label{eq:gap-equation}
\end{align}
where $N_0$ denotes the density of states (DOS) in the normal state on the Fermi level and $U(\alpha)=2\lambda\cos\alpha$ is the coupling potential. $\lambda$ satisfies
\begin{align}
  \frac{1}{N_0\lambda} = \ln\frac{T}{T_{\text{c0}}}+\sum_{n=0,\omega_n\le\omega_{\text{c}}}\frac{1}{n+1/2}
  ,
\end{align}
where $\omega_{\text{c}}$ is the cutoff of the Matsubara frequency and $T_{\text{c0}}$ is the critical temperature without impurities.

\subsection{Solution for bulk }
\label{subsection:bulk}
Here, we summarize the solution for the homogeneous case, which will serve as basic materials to discuss the physical properties of a single vortex. 

For a homogeneous system, $\bm{\nabla}\check{g}$ is zero and $\average{f} = \average{\hconj{f}} = 0$ in a chiral $p$-wave SC, and thus in the bulk, $\Sigma_{12} = \Sigma_{21} = 0$. 
The solution to \eqref{eq:Eilenberger-equation} in the homogeneous case is expressed as
\begin{align}
  \check{g} =-\cpi\frac{\tilde z \check{\tau}_3 -\check{\Delta}}{\sqrt{-\tilde z^2+\abs{\Delta}^2}}.
\label{eq: g-bulk}
\end{align}
Here, $\tilde z=z-\Sigma_{\text{d}}$ is the renormalized frequency, which is determined implicitly by 
\begin{equation}
\tilde{z}-\frac{\cpi\Gamma_{\text{n}} \tilde{z}}{\sqrt{-\tilde z^2+\abs{\Delta}^2}\left(\cos^2\delta_0-\tilde{z}^2(-\tilde z^2+\abs{\Delta}^2)^{-1}\sin^2\delta_0\right)}=z. 
\label{eq: root-for-tilde-z}
\end{equation}
The branch in \eqref{eq: root-for-tilde-z} should be taken so that $\tilde{z}$ is an analytic function for the upper or lower half complex plane of $z$ and satisfies the asymptotics
\begin{align}
\tilde{z} &\to z+\ci\cpi\Gamma_{\text{n}}\sgn(\Im z) ,& \abs{\Im z}\to \infty. 
\end{align}

From the above equations and \eqref{eq:gap-equation}, we can obtain the magnitude of the pair potential in the bulk $\Delta_{\text{b}}(T,\Gamma_{\text{n}},\delta_0)$ under given $\Gamma_{\text{n}}$ and $\delta_0$ at temperature $T$. 

\begin{figure}
  \centering
  \includegraphics[width=0.7\columnwidth,clip]{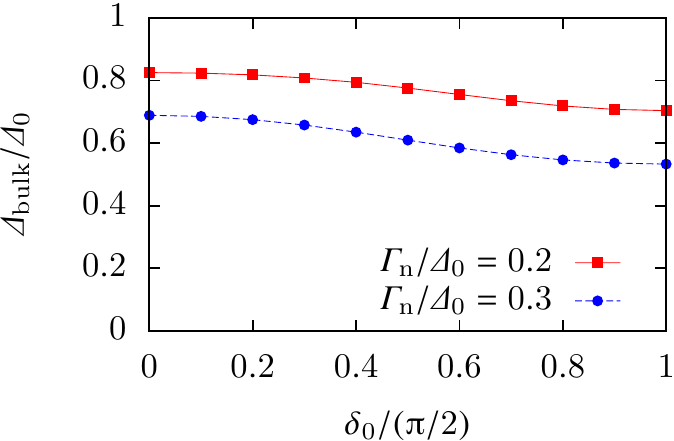}
  \caption{(Color online) Dependence of bulk pair potential $\abs{\Delta_{\text{b}}(T=0.3T_{\text{c0}},\Gamma_{\text{n}})}$ on the phase-shift $\delta_0$ (the parameters of the calculation are the same as in the Sec. \ref{section:method}).}
  \label{fig:pair-potential-vs-phase-shift}
\end{figure}

The critical temperature  $T_{\text{c}}$ under a given $\Gamma_{\text{n}}$ can be derived by linearizing \eqref{eq:gap-equation}; it is given by the Abrikosov-Gor'kov law for anisotropic superconductors\cite{Larkin1965} and does not depend on $\delta_0$\cite{Mineev}. On the other hand, the magnitude of the pair potential $\abs{\Delta_{\text{b}}(T, \Gamma_n)}$ depends on $\delta_0$ (e.g., Fig.~\ref{fig:pair-potential-vs-phase-shift}).

In the following, we see that the existence of the low-energy DOS in the bulk is crucial to the states in the vortex core. For a relatively large phase-shift $\delta_0$, the DOS emerges on the Fermi surface\cite{Maki1999, Maki2000}. The critical phase-shift $\delta_{\text{c}}$ is described as 
\begin{align}
  \delta_{\text{c}}=\arccos\sqrt{\frac{\Gamma_{\text{n}}}{\abs{\Delta_{\text{b}}(\Gamma_{\text{n}};\delta_{\text{c}})}}}
  \label{eq:critical-phase-shift},
\end{align}
as explained in the Appendix~\ref{appendix:critical-phase-shift}. 
The DOS in the bulk can be derived from $-N_0\Im g(z)/\cpi\,|_{z\rightarrow \epsilon+\ci 0}$ with \eqref{eq: g-bulk} and \eqref{eq: root-for-tilde-z}, and is later shown by the red curves in Fig.~\ref{fig:bulk-core-t03g03m}. %

\subsection{Single vortex}
\label{subsection:vortex}
We assume that the vortex is axial symmetric and we take the center of the vortex as the origin of the coordinate system $\bm{r}=(r\cos\phi, r\sin\phi)$. 
To solve \eqref{eq:Eilenberger-equation}, we transform the equation to two independent Riccati-type ordinary differential equations\cite{Nagato1993, Schopohl1995, Schopohl1998},
\begin{align}
  \ci \bm{v}_{\text{F}}\cdot\bm{\nabla}\gamma&=  - (\cconj{\Delta}-\Sigma_{21})\gamma^2 - 2(z - \Sigma_{\text{d}})\gamma - (\Delta + \Sigma_{12})
  \label{eq:Riccati-type-equations-1}
  ,\\
  \ci \bm{v}_{\text{F}}\cdot\bm{\nabla}\bar\gamma&= -(\Delta+\Sigma_{12})\bar\gamma^2+2(z-\Sigma_{\text{d}})\bar\gamma-(\cconj{\Delta}-\Sigma_{21})
  \label{eq:Riccati-type-equations-2}
  ,
\end{align}
and the Green's function is represented by $\gamma$ and $\bar\gamma$ as
\begin{align}
  \check g = \frac{-\ci\cpi\sgn(\Im z)}{1+\gamma\bar \gamma}\begin{pmatrix}
    1-\gamma\bar\gamma & 2\ci\gamma \\
    -2\ci\bar\gamma & -(1-\gamma\bar\gamma)
  \end{pmatrix}
  .
\end{align}
The path along which Eqs.~\eqref{eq:Riccati-type-equations-1} and \eqref{eq:Riccati-type-equations-2} are solved is parallel to the momentum of the quasiparticles and called a quasiclassical trajectory. We denote by $s=r\cos(\phi-\alpha)$ and $b=r\sin(\phi-\alpha)$, respectively, the coordinates along and perpendicular to the quasiclassical trajectory.

We denote the pair potentials for antiparallel and parallel vortices  by $ \Delta^{\text{(a)}}$ and $\Delta^{\text{(p)}}$ respectively, which have the forms \cite{Hayashi2005, Tanuma2009, Sauls2009}
\begin{align}
  \Delta^{\text{(a)}}(\bm{r}, \bm{\hat k}) &= \Delta_{+}^{\text{(a)}}(r)\ce^{\ci(\phi-\alpha)} + \Delta_{-}^{\text{(a)}}(r)\ce^{\ci(-\phi+\alpha)}
  ,\\
  \Delta^{\text{(p)}}(\bm{r}, \bm{\hat k}) &= \Delta_{+}^{\text{(p)}}(r)\ce^{\ci(\phi+\alpha)}+\Delta_{-}^{\text{(p)}}(r)\ce^{\ci(3\phi-\alpha)}.
\end{align}
Here, the subscript ${+}$ (${-}$) indicates the dominant (induced) part of the pair potential; we fix the overall phase of the pair potentials so that $\Delta_{+}^{\text{(a,p)}}(r\rightarrow \infty)$ is positive real. 

Each matrix element of $\check\Sigma$ is coupled to $z$, $\Delta$, and $\cconj{\Delta}$ as in \eqref{eq:Eilenberger-equation}, and the rotational symmetry constrains them to have the following forms\cite{Hayashi2005}:
\begin{align}
  \Sigma_{\text{d}}(r,\phi+\theta) &= \Sigma_{\text{d}}(r,\phi) = \Sigma_{\text{d}}(r) \label{eq:Sigma-diagonal},\\
  \Sigma_{12}(r,\phi+\theta) &= \Sigma_{12}(r,\phi)\ce^{+\ci l\theta} \label{eq:Sigma-off-diagonal-1},\\
  \Sigma_{21}(r,\phi+\theta) &= \Sigma_{21}(r,\phi)\ce^{-\ci l\theta} \label{eq:Sigma-off-diagonal-2}
  .
\end{align}
Here, $l$ is the total angular momentum of the system: $l=0$ for the antiparallel vortices and $l=2$ for the parallel vortices. 
For both cases, we assume that $\gamma$ and $\bar\gamma$ satisfy
\begin{align}
  \gamma(s,b) &= \frac{-\tilde z +\sgn(\Im z)\ci\sqrt{-\tilde z^2+\abs{\Delta}^2}}{\cconj{\Delta}} & \text{for } s&=-\sgn(\Im z)s_{\text{c}}, \\
  \bar\gamma(s,b) &= \frac{\tilde z-\sgn(\Im z)\ci\sqrt{-\tilde z^2+\abs{\Delta}^2}}{\Delta}& \text{for } s&=+\sgn(\Im z)s_{\text{c}}
\end{align}
at a sufficiently large cutoff $s_{\text{c}}\gg\xi_0$.

\subsection{Kramer-Pesch approximation and its variant}
\label{subsection: KPA}
As a prerequisite to discuss the results of numerical calculations, we summarize an approximate expression for the quasiclassical Green's function near the vortex cores and its variant. 
In pure superconductors $(\check\Sigma=0)$, the Green's function near the vortex core at a low energy, $|\epsilon|\ll |\Delta_{\text{b}}|$, is known to be described by the analytical expression \cite{Kato2000}
\begin{align}
  \check g^{\text{(a,p)}}(\epsilon,{\bm r},\alpha) = \frac{\cpi v_{\text{F}}\exp\left[-u^{\text{(a,p)}}(s;b)\right]}{2C^{\text{(a,p)}}(b)}\frac{\check M^{\text{(a,p)}}(\alpha)}{\epsilon-E_{\Delta}^{\text{(a,p)}}(b)+\ci\eta}
  \label{eq:perturbation-vortex-green-pure}
  .
\end{align}
The quantities in this expression denote
\begin{widetext}
\begin{align}
  u^{\text{(a)}}(s;b) &= \frac{2}{v_{\text{F}}}\int_{0}^{|s|}\dd s'\frac{s'\left(\Delta^{\text{(a)}}_+(\sqrt{s'^2 +b^2})+\Delta^{\text{(a)}}_-(\sqrt{s'^2 +b^2}) \right)}{\sqrt{s'^2 +b^2}}
  ,\\
  u^{\text{(p)}}(s;b) &= \frac{2}{v_{\text{F}}}\int_{0}^{|s|}\dd s'
\left(\frac{s'\Delta^{\text{(p)}}_+ (\sqrt{s'^2 +b^2})}{\sqrt{s'^2 +b^2}}
+\frac{(s'^3 -3b^2 s')\Delta^{\text{(p)}}_-(\sqrt{s'^2 +b^2})}{(s'^2 +b^2)^\frac32}\right)
  ,\\
  C^{\text{(a,p)}}(b)&= \int_0^\infty\dd s' \exp[-u^{\text{(a,p)}}(s';b)]
  ,\\
E_{\Delta}^{(\text{a})}(b)&= \frac{1}{C^{(\text{a})}(b)}\int_{0}^{\infty}\dd s'
\frac{b\left(\Delta^{\text{(a)}}_+(\sqrt{s'^2 +b^2})-\Delta^{\text{(a)}}_-(\sqrt{s'^2 +b^2}) \right)}{\sqrt{s'^2 +b^2}}\exp[-u^{\text{(a)}}(s';b)]
  ,\\
E_{\Delta}^{(\text{p})}(b)&= \frac{1}{C^{(\text{p})}(b)}
\int_{0}^{\infty}\dd s'
\left(\frac{b\Delta^{\text{(p)}}_+(\sqrt{s'^2 +b^2})}{\sqrt{s'^2 +b^2}}+\frac{(3bs'^2 -b^3)\Delta^{\text{(p)}}_-(\sqrt{s'^2 +b^2})}{(s'^2 +b^2)^\frac32}\right)\exp[-u^{\text{(p)}}(s';b)]
  ,\\
  \check M^{(\text{a})}(\alpha) &=
  \begin{pmatrix} 1 & -\ci\\ -\ci & -1\end{pmatrix},
\quad
  \check M^{(\text{p})}(\alpha)=
  \begin{pmatrix} 1 & -\ci\ce^{2\ci \alpha}\\ -\ci\ce^{-2\ci \alpha} & -1\end{pmatrix}
    .
\end{align}
\end{widetext}
In impure but clean superconductors, we expect that an expression similar to 
 \eqref{eq:perturbation-vortex-green-pure} gives a good approximation to $\check{g}$\cite{Kato2000}:
\begin{align}
  \check g^{\text{(a,p)}}(\epsilon,{\bm r},\alpha)
  = \frac{\cpi v_{\text{F}}\exp\left[-u^{\text{(a,p)}}(s;b)\right]}{2C^{\text{(a,p)}}(b)}\frac{\check M^{\text{(a,p)}}(\alpha)}{\epsilon-E_{\Delta}^{\text{(a,p)}}(b)-E_{\Sigma}^{\text{(a,p)}}(\epsilon,b,\alpha)}
+\check g^{\text{(a,p)}}_{\text{cont}}.
\label{eq:perturbation-vortex-green-impure}
\end{align}
The first and second terms in the right-hand side represent, respectively, the contributions from the vortex-ABS and the impurity bands in the bulk. 
Here,
\begin{align}
  & E_{\Sigma}^{(\text{a,p})}(\epsilon,b,\alpha)
  = \frac{1}{C^{(\text{a,p})}(b)}
\int_{0}^{\infty}\dd s\tilde{\Sigma}^{(\text{a,p})}(\epsilon,s,b,\alpha)\exp[-u^{\text{(a,p)}}(s;b)]
\label{eq:esigma-from-tilde-sigma-by-integration}
\end{align}
with 
\begin{align}
  \tilde\Sigma^{(\text{a})}(\epsilon,s,b,\alpha)
  &= \Sigma_{\text{d}}(\epsilon,s,b) -\frac{\ci}{2}\left(\Sigma_{12}(\epsilon,s,b) +\Sigma_{21}(\epsilon,s,b)\right)
,\\
  \tilde\Sigma^{(\text{p})}(\epsilon,s,b,\alpha)
  &= \Sigma_{\text{d}}(\epsilon,s,b) -\frac{\ci}{2}\left(\Sigma_{12}(\epsilon,s,b)\ce^{-2\ci \alpha} +\Sigma_{21}(\epsilon,s,b)\ce^{+2\ci \alpha}\right)
\label{eq:effective-sigma-in-core}
\end{align}
represents the self-energy effect on the excitation spectrum; the real and imaginary parts represent, respectively, the renormalization of the spectrum and damping effects. The spectrum of the vortex-ABS is determined implicitly by
\begin{align}
\epsilon-E^{(\text{a,p})}_\Delta (b)-\Re E^{(\text{a,p})}_\Sigma (\epsilon,b)
&\sim \epsilon-E^{(\text{a,p})}_\Delta (b)-\Re E^{(\text{a,p})}_\Sigma (0,b)-\epsilon\frac{\partial}{\partial\epsilon'}\Re E^{(\text{a,p})}_\Sigma (\epsilon',b)\Big|_{\epsilon'=0}
\nonumber\\
&=0,
\end{align}
which yields the excitation energy of the vortex-ABS
\begin{equation}
\epsilon=Z(b)\left(E^{(\text{a,p})}_\Delta (b)-\Re E^{(\text{a,p})}_\Sigma (0,b)\right)\equiv\tilde{E}^{(\text{a,p})}(b)
\end{equation}
with 
\begin{equation}
Z=\left(1-\frac{\partial}{\partial\epsilon'}\Re E^{(\text{a,p})}_\Sigma (\epsilon',b)\Big|_{\epsilon'=0}\right)^{-1}.
\end{equation}
Here, $Z$ is the renormalization factor of the excitation spectrum. Near $\epsilon=\tilde{E}(b)$, \eqref{eq:perturbation-vortex-green-impure} reduces to 
\begin{align}
  \check g^{\text{(a,p)}}(\epsilon,{\bm r},\alpha)
  &\sim\frac{\cpi v_{\text{F}}\exp\left[-u^{\text{(a,p)}}(s;b)\right]}{2C^{\text{(a,p)}}(b)}
  \frac{Z^{\text(a,p)}(b)\check M^{\text{(a,p)}}(\alpha)}{\epsilon-\tilde{E}^{\text{(a,p)}}(b)+\ci\Gamma^{(\text{a,p})}(\epsilon,b)}
+\check g^{\text{(a,p)}}_{\text{cont}}
  \label{eq:perturbation-vortex-green-impure-reduction}
\end{align}
with
\begin{align}
  \Gamma^{(\text{a,p})}(\epsilon, b)\equiv -Z(b)\Im E^{(\text{a,p})}_{\Sigma}(\epsilon,b).
  \label{eq:gamma-from-esigma-and-z}
\end{align}
{
When the peak broadening by impurities is not so large, the function form of \eqref{eq:perturbation-vortex-green-impure-reduction} tells us that $\Gamma^{\text{(a,p)}}(\epsilon=\tilde E(b), b)$ corresponds to the inverse lifetime of the bound states.

\subsection{Method for numerical calculation}
\label{section:method}
We use the adaptive fourth- and fifth-order Runge-Kutta method\cite{Shampine1986} to solve \eqref{eq:Riccati-type-equations-1} and \eqref{eq:Riccati-type-equations-2}. The discrete sampling points $r_i$ along the radial line are taken as $r_i = \tilde r(\exp(R_i)-1)$ as in Ref.~\citen{Hayashi2013-swave}. For each sampling point, we take equally spaced trajectories in the momentum space. The numbers of trajectories are 200--3200, which are set to be large enough to accomplish the accuracy and depend on the given parameters. We first set an initial value for the energy gap to solve the Riccati equations and obtain the quasiclassical Green's function for all Matsubara frequencies with $\tilde r=0.01\xi_0$ and $R_i=0.07i$, then solve gap equation \eqref{eq:gap-equation}, and repeat these steps until the maximum relative or absolute difference of $\Delta$ (for Matsubara frequencies) or $\average{\check g(z)}$ (for real frequencies) is less than $10^{-3}$ for all $\bm{r}$, as described in Refs.~\citen{Kato2002, Hayashi2005, Tanuma2009, Kurosawa2014}. We take the cutoffs $\omega_{\text{c}} = 10\Delta_0$ and $x_{\text{c}}=100\cpi\xi_0$. After calculating the order parameter self-consistently, we calculate Green's functions for real-time frequencies $\epsilon+\ci\eta$ by solving \eqref{eq:Riccati-type-equations-1} and \eqref{eq:Riccati-type-equations-2}, with $\tilde r=0.001\xi_0$ and $R_i=0.06i$. We first set the smearing factor as $\eta=0.005\Delta_0$ and reduce it to zero in the final results presented in this paper. We use the extended Anderson mixing scheme\cite{Eyert1996} to accelerate the convergence.

\section{Numerical Results}
\label{section:numerical-results}

\begin{figure}
  \centering
  \includegraphics[width=0.45\textwidth]{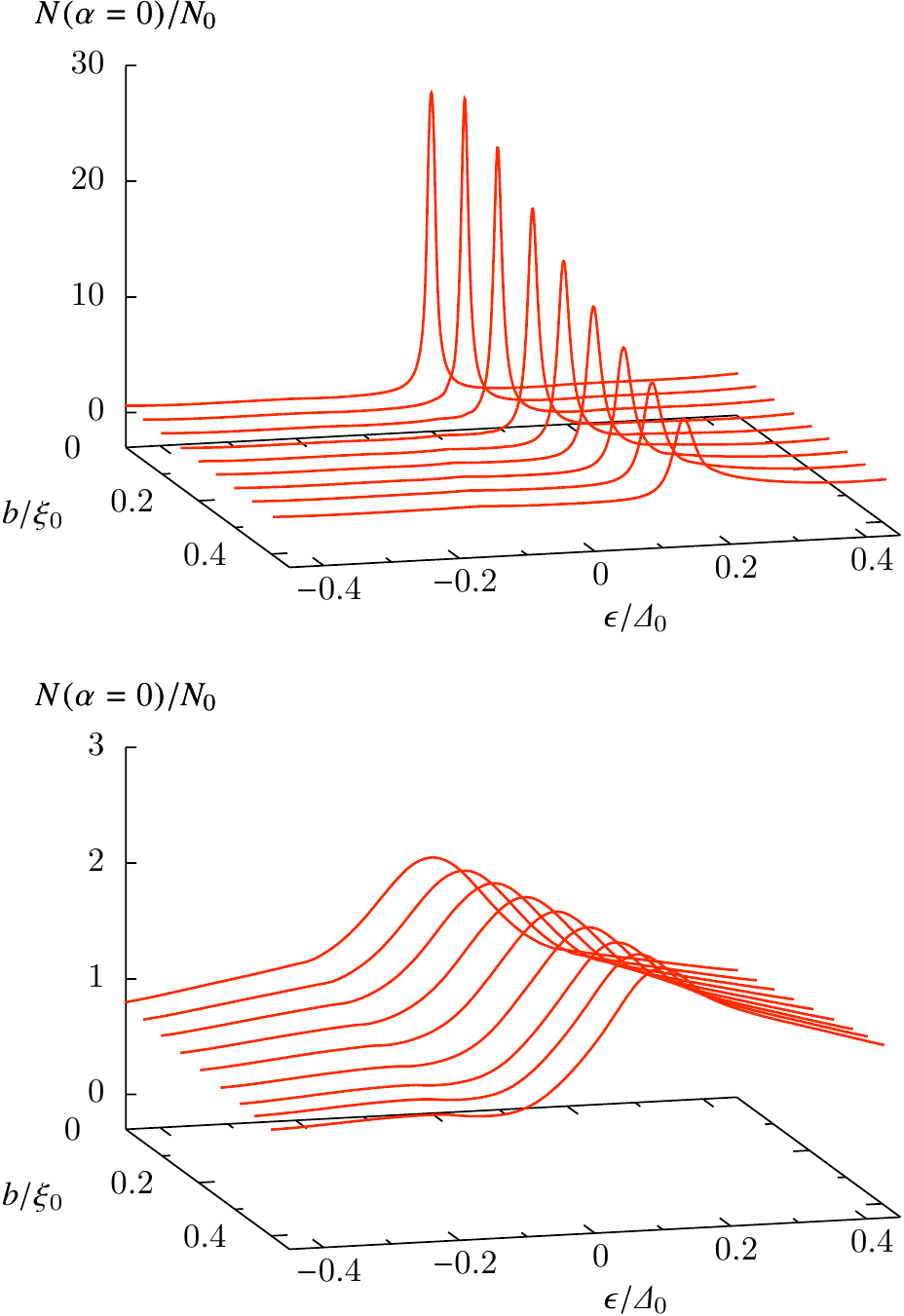}
  \caption{(Color online) Angular resolved LDOS for $T=0.3T_{\text{c0}}$, $\Gamma_{\text{n}}=0.3\Delta_0$, and $\delta_0=0$ (Born limit). Top:  $l_z=0$ (antiparallel); bottom:  $l_z=2$ (parallel).}
  \label{fig:angular-resolved-shape-Born}
\end{figure}

\begin{figure}
  \centering
  \includegraphics[width=0.45\textwidth]{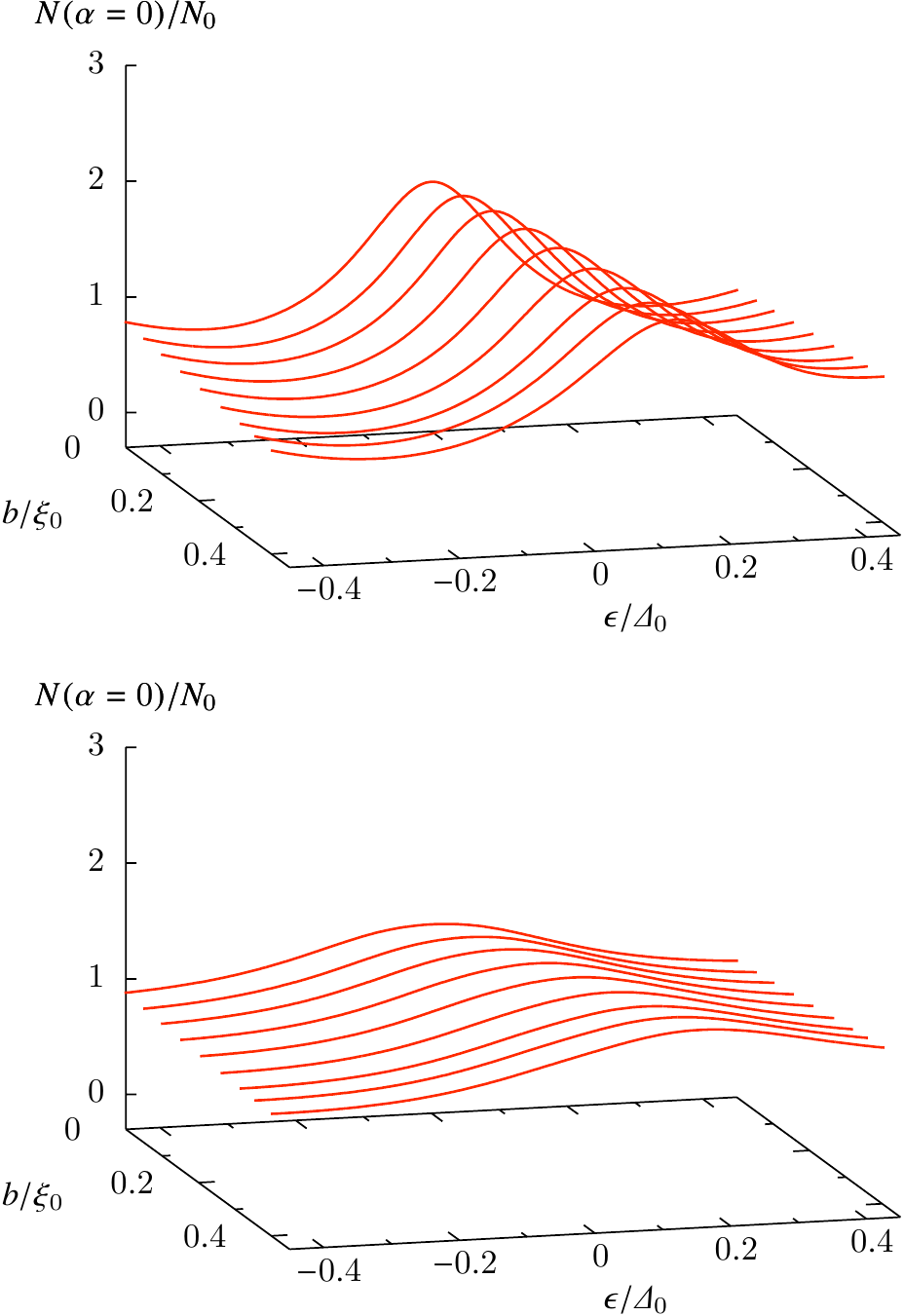}
  \caption{(Color online) Angular resolved LDOS for $T=0.3T_{\text{c0}}$, $\Gamma_{\text{n}}=0.3\Delta_0$, and $\delta_0=\cpi/2$ (unitary limit). Top:  $l_z=0$ (antiparallel); bottom:  $l_z=2$ (parallel).}
  \label{fig:angular-resolved-shape-unitary}
\end{figure}

\begin{figure}
  \centering
  \includegraphics[width=0.24\textwidth]{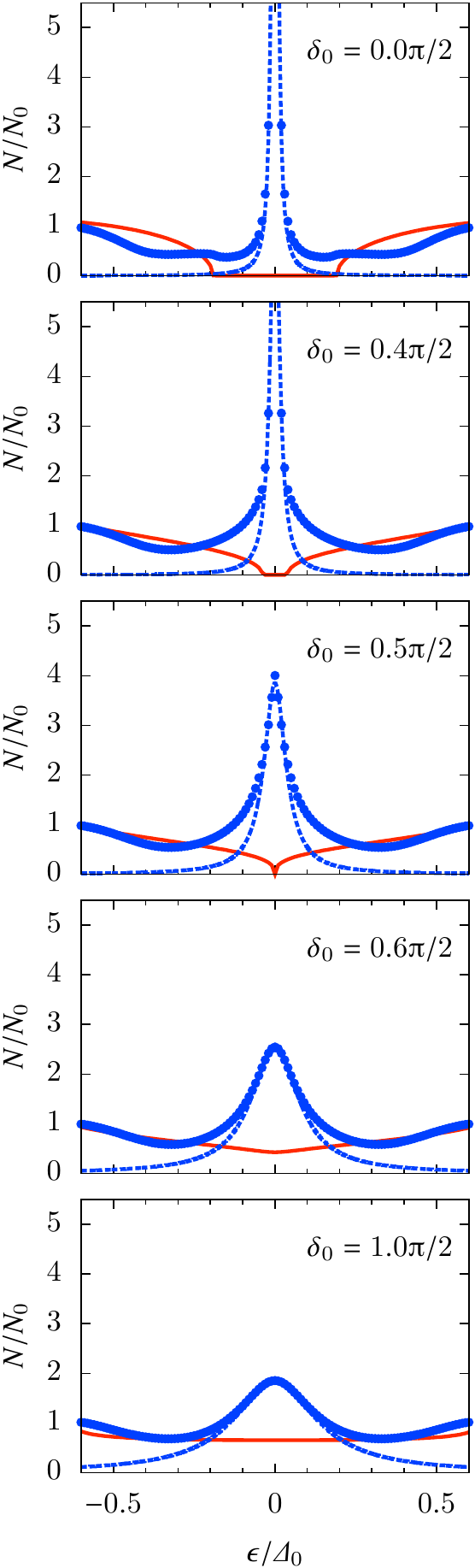}
  \includegraphics[width=0.24\textwidth]{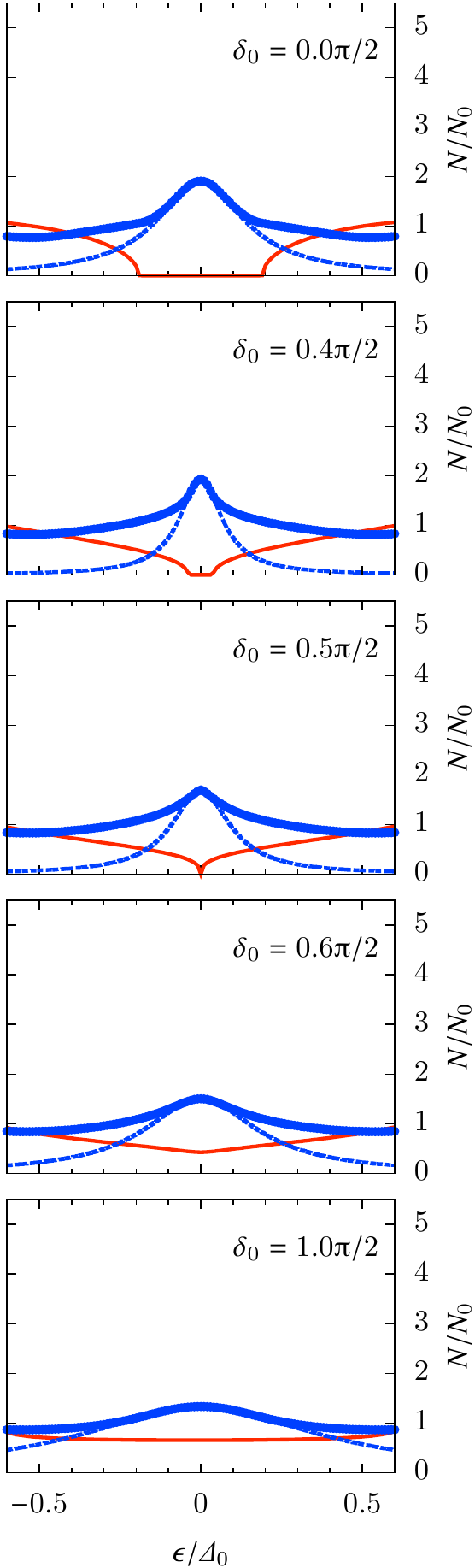}
  \caption{(Color online) Angular resolved LDOS of the bulk and the core at $T=0.3T_{\text{c0}}$, $\Gamma_{\text{n}}=0.3\Delta_0$. For this parameter, $\delta_{\text{c}}$ is about $0.52\cpi/2$. Blue circles: angular resolved LDOS at the core; Blue dashed curve: fitted Lorentzian function; red solid curve: angular resolved LDOS at the bulk. Left: antiparallel vortex ($l_z=0$); right: parallel vortex ($l_z=2$).}
  \label{fig:bulk-core-t03g03m}
\end{figure}

\begin{figure}
  \centering
  \includegraphics[width=0.49\textwidth]{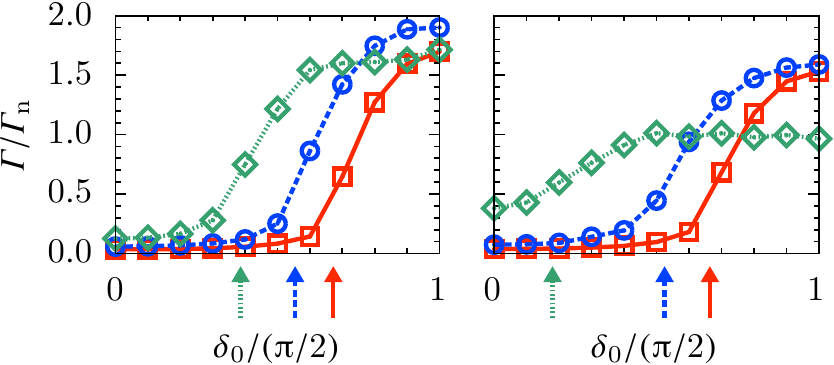}
  \caption{(Color online) Phase-shift dependence of the half width of the zero-energy angular resolved LDOS at the center of the antiparallel vortex. The arrows indicate the impurity phase-shift where the minimum excitation energy becomes zero in the bulk under given $T$ and $\Gamma_{\text{n}}$ in the bulk (each arrow corresponds to a parameter with the same color and line type). Left:  $T=0.1 T_{\text{c}}$; right:  $T=0.3 T_{\text{c}}$; red squares and solid line:  $\Gamma_{\text{n}}=0.2\Delta_0$; blue circles and dashed line:  $\Gamma_{\text{n}}=0.3\Delta_0$; green diamonds and dotted line:  $\Gamma_{\text{n}}=0.4\Delta_0$. }
\label{fig:angular-resolved-width-antiparallel}
\end{figure}

\begin{figure}
  \centering
  \includegraphics[width=0.49\textwidth]{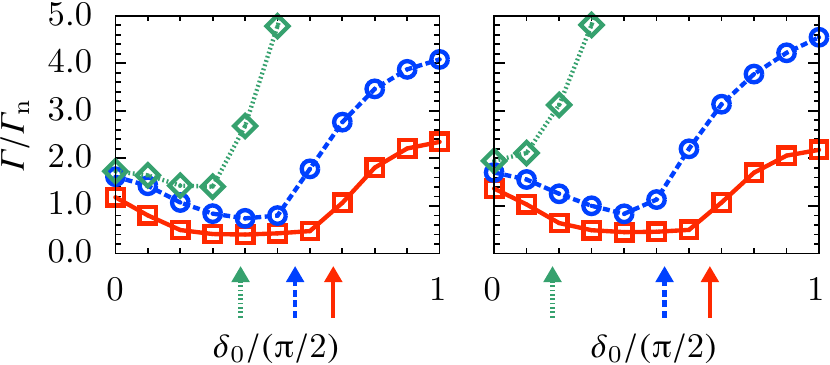}
  \caption{(Color online) Phase-shift dependence of the half-width of the zero energy angular resolved LDOS at the center of the parallel vortex. The arrows are the same as these in Fig.~\ref{fig:angular-resolved-width-antiparallel}. Left:  $T=0.1 T_{\text{c}}$; Right:  $T=0.3 T_{\text{c}}$; red squares and solid line:  $\Gamma_{\text{n}}=0.2\Delta_0$; blue circles and dashed line:  $\Gamma_{\text{n}}=0.3\Delta_0$; green diamonds and dotted line:  $\Gamma_{\text{n}}=0.4\Delta_0$.}
  \label{fig:angular-resolved-width-parallel}
\end{figure}

Figure~\ref{fig:angular-resolved-shape-Born} shows the angular resolved LDOS $N(\epsilon,\bm{r}) = -N_0\Im g(\epsilon,{\bm r},\alpha)/\cpi$ at $s=0$ for various values of $b$ near the vortex core in the presence of Born scatterers (i.e., $\delta_0=0$). The angular resolved LDOS does not depend on $\alpha$.  We see that there is a single peak for each curve in the energy range shown here. The peak position $\epsilon=\epsilon(b)$ shifts to a higher energy as the impact parameter $b$ increases and we identify $\epsilon(b)$ with the excitation energy of the vortex-ABS. At the vortex center $\bm{r}=\bm{0}$, the angular resolved LDOS is equivalent to the LDOS\cite{Kurosawa2014}. 
We see a clear difference in the spectrum between the antiparallel (top) and parallel vortices (bottom). The sharper peaks at the top of Fig.~\ref{fig:angular-resolved-shape-Born} imply that the vortex-ABS within the antiparallel vortex is more tolerant toward the Born scatterers than that within the parallel vortex. 

In the unitary limit, in contrast, the spectra are more broadened than those in the Born limit, both in the antiparallel and parallel vortices (Fig. \ref{fig:angular-resolved-shape-unitary}). We thus see that the unitary impurities affect the vortex-ABS much more than the Born impurities. 

To see the $\delta_0$-dependence of the spectral shape, we focus on the angular resolved LDOS for various values of $\delta_0$ at the vortex core ($s=b=0$), which is shown in Fig.~\ref{fig:bulk-core-t03g03m} (left column for antiparallel vortex and right column for parallel vortex) by blue dots. For comparison, we show the DOS of the impurity band in the bulk by the red curves. We note that $\delta_{\text{c}}\approx 0.52\cpi/2$ for the present parameters $T=0.3T_{\text{c}}$ and $\Gamma_{\text{n}}=0.3\Delta_0$. 

We see in the left column (antiparallel vortex) that the peak in the spectrum is sharp for $\delta_0=0$ and $0.2\cpi$, while the peak becomes broader as $\delta_0(>0.25\cpi)$ increases. In the right column (parallel vortex) in Fig.~\ref{fig:bulk-core-t03g03m}, the spectrum becomes narrower and then becomes broader as $\delta_0$ increases from $\delta_0=0$ to $0.25\cpi$. When $\delta_0\gtrsim 0.25\cpi$, the spectrum becomes broader as $\delta_0$ increases.

To quantify the effect of impurities, we fit the Lorentzian function $c_1/(\epsilon^2+(\Gamma/\cpi)^2)$ via some constant $c_1$ and the half width $\Gamma$ of the angular resolved LDOS at the core to the calculated value of the angular resolved LDOS at $\epsilon/\Delta_0=0$, $\pm0.005$, $\pm0.010$ and $\pm0.015$. The resultant Lorentzian functions are plotted by blue-dotted curves in Fig.~\ref{fig:bulk-core-t03g03m}. The obtained $\Gamma$ are plotted in Fig.~\ref{fig:angular-resolved-width-antiparallel} for the antiparallel vortex and in Fig.~\ref{fig:angular-resolved-width-parallel} for the parallel vortex. The arrows in these figures indicate  $\delta_{\text{c}}$ for each parameter. 

For the antiparallel vortices (Fig. \ref{fig:angular-resolved-width-antiparallel}) at $T=0.1 T_{\text{c}}$ (a) and $T=0.3 T_{\text{c}}$ (b), we see that the half width of the spectral peak is small for $\delta_0 < \delta_{\text{c}}$, while it becomes larger as $\delta_0(>\delta_{\text{c}})$ increases for $\Gamma_{\text{n}}=0.2\Delta_0$ and $0.3\Delta_0$. The crossover from small $\Gamma$ to large $\Gamma$ around $\delta_0=\delta_{\text{c}}$ is more gradual at $\Gamma_{\text{n}}=0.4 \Delta_0$.
As a counter-intuitive result in Fig.~\ref{fig:angular-resolved-width-antiparallel}, we note that the peak width $\Gamma$ at $\Gamma_{\text{n}}=0.3\Delta_0$ is larger than the that at $\Gamma_{\text{n}}=0.4\Delta_0$ near the unitary limit, yet the weight of the peak at $\Gamma_{\text{n}}=0.3\Delta_0$ is larger than that at $\Gamma_{\text{n}}=0.4\Delta_0$. However, this just implies that the fittings are not quantitatively good in these parameters and has no significant physical meaning.
The corresponding data of the angular resolved LDOS are in Appendix \ref{appendix:supplementary}.

For the parallel vortices (Fig.~\ref{fig:angular-resolved-width-parallel}) at $T=0.1 T_{\text{c}}$ (a) and $T=0.3 T_{\text{c}}$ (b), we see that the half width of the spectral peak depends on $\delta_0$ nonmonotonically when $\delta_0 < \delta_{\text{c}}$ while it increases monotonically as $\delta_0(>\delta_{\text{c}})$ increases for $\Gamma_{\text{n}}=0.2\Delta_0$ and $0.3\Delta_0$. This behavior is consistent with what we have observed in Fig.~\ref{fig:bulk-core-t03g03m}. We note that the results for $\Gamma_{\text{n}}=0.4 \Delta_0$ in Fig.~\ref{fig:angular-resolved-width-parallel} deviate from this behavior. This value of $\Gamma_{\text{n}}$ might be too large to discuss the systematic properties of the impurity effect in clean \(p\)-wave superconductors. 

The spectral broadening for $\delta_0 \in [\delta_{\text{c}},\cpi/2]$ inherently requires the self-consistent treatment of the scattering process. We show an example. We numerically find the scaling law 
\begin{align}
\Gamma &\propto {\Gamma_{\text{n}}}^{\vartheta},& \vartheta\approx \frac34\quad\mbox{for}\quad\delta_0=\frac{\cpi}{2} 
\label{eq:43}
\end{align}
for the antiparallel vortex in clean superconductors $\Gamma_{\text{n}}/\Delta_0 \ll 0.1$ at temperatures $T/T_{\text{c}}=0.2$ and $0.3$ (see Fig.~\ref{fig:unitary-ZEDOS-gamma_n-dependency}). We do not find the explanation for this scaling relationship through the non-self-consistent $t$-matrix approximation.

\begin{figure}
  \centering
  \includegraphics[width=0.4\textwidth]{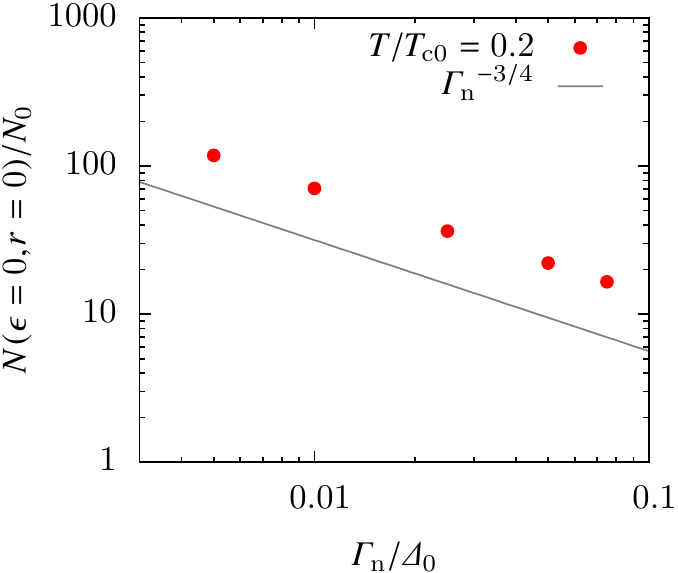}\hspace{0.1\textwidth}
  \includegraphics[width=0.4\textwidth]{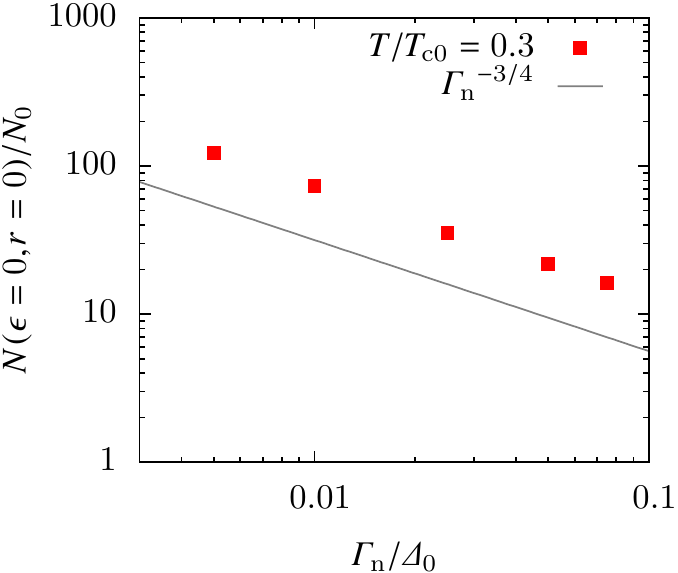}
  \caption{(Color online) Zero-energy (angular resolved) LDOS at the center of the vortex $N(\epsilon=0,r=0)/N_0$ with $\delta_0=\cpi/2$ (left: $T/T_{\text{c0}}=0.2$; right: $T/T_{\text{c0}}=0.3$). The powers obtained by least-squares fitting are $-0.724\pm0.002$ (left) and $-0.747\pm0.008$ (right). The scattering rate $\Gamma$ itself is hardly quantified via numerical calculation because of the very sharp  spectrum with low impurities; therefore, we assume that the shape of the spectrum is Lorentzian. Under this assumption, the inverse of the magnitude of the LDOS is proportional to the scattering rate $\Gamma$.
  }
  \label{fig:unitary-ZEDOS-gamma_n-dependency}
\end{figure}

Now we discuss the role of $\delta_0$ on the basis of \eqref{eq:esigma-from-tilde-sigma-by-integration} with the numerically obtained $\tilde{\Sigma}$ and $C(b)$ as the input data.
Equation \eqref{eq:esigma-from-tilde-sigma-by-integration} implies that $\Gamma$ is given by the average of $\tilde{\Sigma}$ with the weight ${\ce}^{-u}/C(b)$ along the trajectory with a constant $b$. $\tilde{\Sigma}$ can thus be regarded as the local scattering rate.

Figure \ref{fig:effective-Sigma} shows that $\tilde\Sigma$ vanishes practically for the antiparallel vortex for $\delta_0<\delta_{\text{c}}$. This implies that $\delta_0$ does not affect the spectrum as long as $\delta_0<\delta_{\text{c}}$.
For the parallel vortex, on the other hand, the local scattering rate near the core is not suppressed in the Born limit. Besides, unlike antiparallel vortices, the suppression of the scattering can be seen at the finite $\delta_0$.

  Figure~\ref{fig:effective-Sigma-weighten} shows how $\tilde\Sigma$ affects the scattering rate at the core of the vortex [the integrated value is proportional to $\Gamma(b=0)$]. For $\delta_0>\delta_{\text{c}}$, we see that the most affected part of $\tilde\Sigma$ is not inside the core. This implies that DOS in the bulk (the so-called ``impurity band'') is responsible for the large value of $\tilde{\Sigma}$ in the bulk (see Fig.~\ref{fig:ldos-r-dependence}) and yields the large value of $\Gamma$ inside the vortex.
We also see that in both cases the impurity effect (contribution to \(\Gamma\)) is suppressed around the vortex cores, in comparison with in the bulk region. 
We thus see that the finite $\delta_0$ affects the bound states in two ways: it suppresses the local scattering rate near the core for $0<\delta_0<\cpi/2$, and generates the DOS and local scattering rate in the whole region on the trajectory for \(\delta_{\text{c}} < \delta < \cpi/2\) (Fig.~\ref{fig:effective-Sigma-weighten}). 

  From the above observation, we may interpret the dependence of $\Gamma$ on $\delta_0$ as follows. For sufficiently large $\delta_0$, the scattering is suppressed in the vicinity of both the antiparallel and parallel vortex cores; in other words, the vortex-ABS, which is localized around the vortex core, is insensitive to the impurities. Besides, the impurities affect the vortex-ABS via the extended states in the bulk region (i.e., the impurity band). It is necessary to consider the mixing of the vortex-ABS and extended states to understand the effects of impurities on the scattering rate of the vortex-ABS.

\begin{figure}
  \centering
  \includegraphics[width=0.4\textwidth]{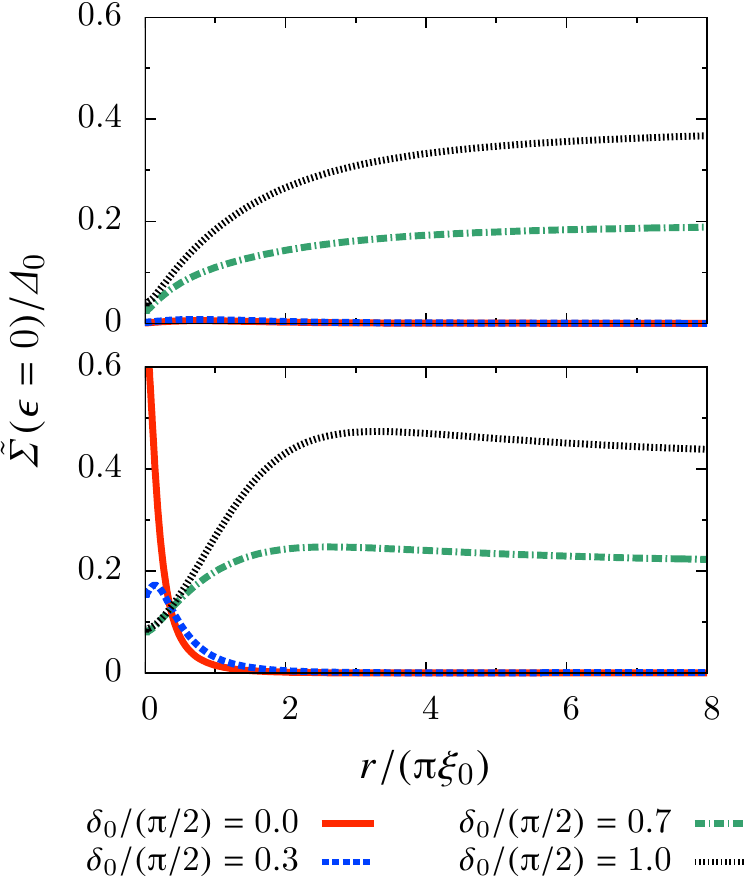}
  \caption{(Color online) Position dependence of $\tilde\Sigma(r,\epsilon=0)$ calculated with the numerically obtained $\Delta$ and $\check\Sigma$ for $T=0.3 T_{\text{c}}$ and $\Gamma_{\text{n}}=0.2\Delta_0$. Top: antiparallel vortex; bottom: parallel vortex.}
  \label{fig:effective-Sigma}
\end{figure}

\begin{figure}
  \centering
  \includegraphics[width=0.4\textwidth]{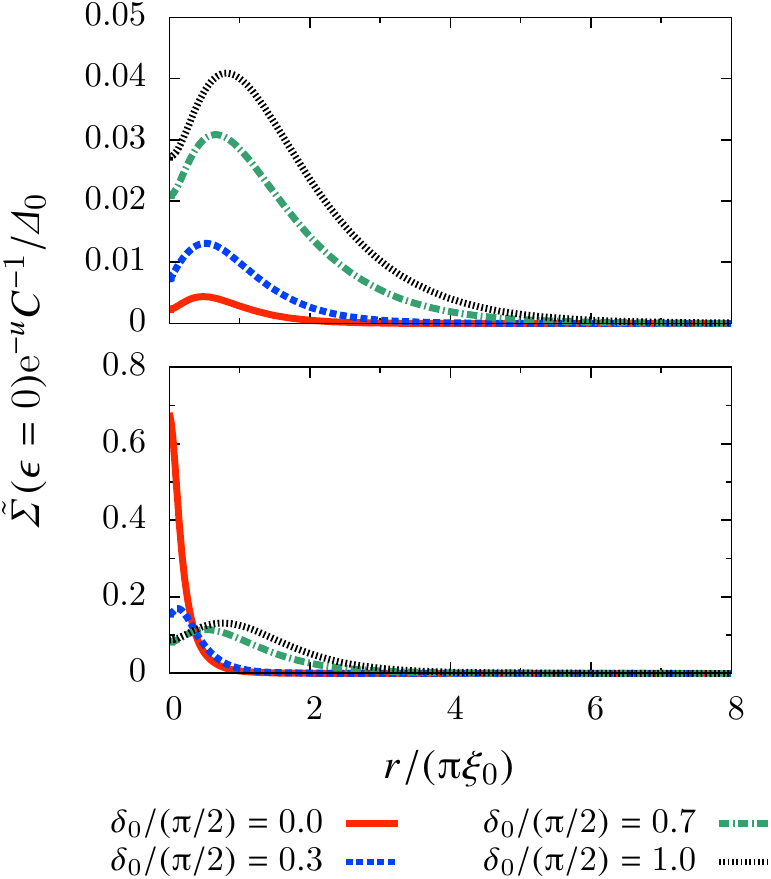}
  \caption{(Color online) Position dependence of $\tilde\Sigma(r,\epsilon=0)\ce^{-u}/C$ of antiparallel vortex for $T=0.3 T_{\text{c}}$ and $\Gamma_{\text{n}}=0.2\Delta_0$. The weight $\ce^{-u}/C$ is calculated with the numerical results, as well as $\Delta$ and $\check\Sigma$. Top: antiparallel vortex; bottom: parallel vortex.}
  \label{fig:effective-Sigma-weighten}
\end{figure}

\begin{figure}
  \centering
  \includegraphics[width=0.4\textwidth]{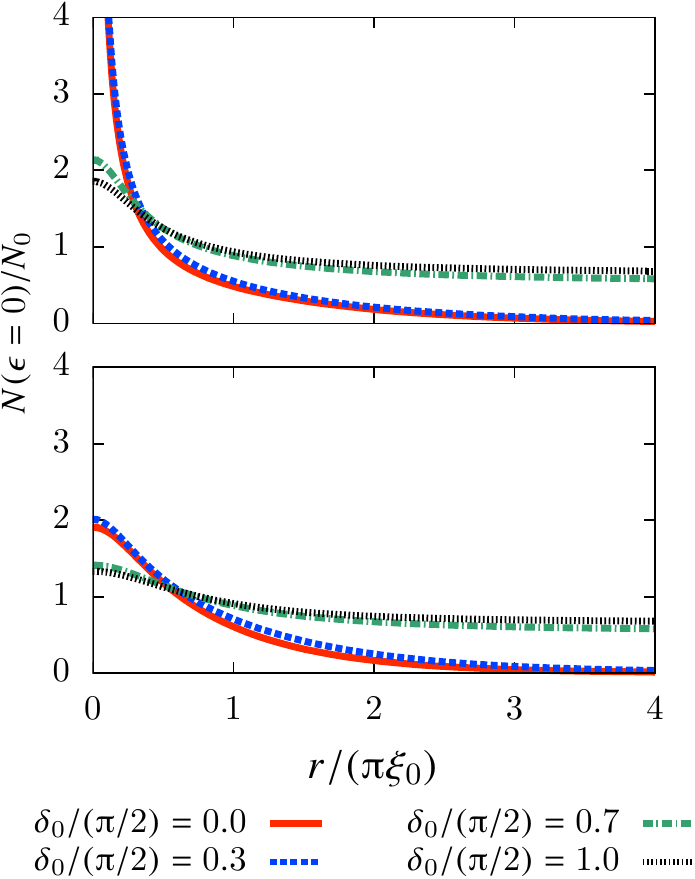}
  \caption{(Color online) Position dependence of zero-energy LDOS of antiparallel vortex for $T=0.3 T_{\text{c}}$ and $\Gamma_{\text{n}}=0.3\Delta_0$. Top: antiparallel vortex; bottom: parallel vortex.}
  \label{fig:ldos-r-dependence}
\end{figure}

We now summarize our findings in this section. 
\begin{enumerate}[(i)]
\item
The spectral shapes at low energy near the vortex cores are clearly different between antiparallel and parallel vortices, particularly from \(\delta_0=0\) (Born limit) to \(\delta_{\text{c}}\).
\item
The spectral shape at low energy near the core of the antiparallel vortex is sharp and the impurities do not have a strong effect for $\delta_0 \in [0,\delta_{\text{c}}]$.
\item
The peak width in the spectral shape at low energy near the core of the parallel vortex depends nonmonotonically on $\delta_0 \in [0,\delta_{\text{c}}]$. The finite $\delta_0$ suppresses the local scattering rate around the core.  
\item
For $\delta_0 \in [\delta_{\text{c}},\cpi/2]$, the spectral shapes become broader as $\delta_0$ increases for antiparallel and parallel vortices. This broadening originates from the impurity band in the bulk. 
\item
In the unitary limit with very small $\Gamma_{\text{n}}$, $\Gamma$ for antiparallel vortices seems to obey a scaling law.
\end{enumerate}
We will discuss (i)-(iii) among the findings in the next section.

\section{Discussion}
%
%
%
%
%
%
%
%
\subsection{Difference in spectral broadening between antiparallel and parallel vortices}
We have seen that spectral broadening is clearly different between antiparallel and parallel vortices, particularly for $\delta_0 \in [0,\delta_{\text{c}}]$. At  the Born limit, earlier studies\cite{Kato2000,Kato2002,Tanuma2009} have reached the same conclusion but the authors of Ref.~\citen{Sauls2009} came to a different conclusion. 
In Ref.~\citen{Sauls2009}, as an origin of this discrepancy between Refs.~\citen{Kato2002,Tanuma2009} and Ref.~\citen{Sauls2009}, the importance of the self-consistent treatment of a two-component pair potential has been pointed out. In Ref.~\citen{Tanuma2009} and the present study, however, calculations have been performed self-consistently. 

One possible source of this discrepancy is the axial symmetric feature\cite{Sauls2013}. An analysis based on the Ginzburg-Landau (GL) theory has been reported for this system\cite{Tokuyasu1990}, where it was argued that the axial symmetry of the order parameter around the vortex can be spontaneously broken for strong-coupling regimes.
We use the quasiclassical theory in the weak-coupling regime\cite{Sauls2009}, and hence near the critical temperature, the vortex should be circular. However, our analysis is carried out at a moderately low temperature, where GL theory is not applicable in general. It is still unclear whether the non-axial symmetric vortex emerges at low temperatures. Further research on the structure of the vortex in the chiral $p$-wave superconductors is desirable.

\subsection{Impurity effect in antiparallel vortex for $\delta_0\in [0,\delta_{\text{c}}]$}\label{sec: antparallel}
In this subsection, we discuss (ii) in our findings with the use of a variant of the Kramer-Pesch approximation summarized in Sec.~\ref{subsection: KPA}. When $\delta\in\LeftClosedRightOpenInterval{0}{\delta_{\text{c}}}$, the impurity band has the energy gap $\epsilon_{\text{ib}}$, the contribution from $\check{g}_{\text{cont}}$ to $\check{g}$ at energy lower than $\epsilon_{\text{ib}}$ might be negligible. With the assumption $\check{g}_{\text{cont}}\sim 0$ and the approximate expression \eqref{eq:perturbation-vortex-green-impure}
, $\Gamma(b)=0$ follows as explained below. 
The self-energy depends on the angular-averaged quasiclassical Green's function. 
In the antiparallel vortex core ,
$\check M^{\text{(a)}}$ does not depend on $\alpha$; the averaged quasiclassical Green's function thus has the form
\begin{align}
  \average{\check g(\epsilon, {\bm r},\alpha)} &=\average{g(\epsilon, {\bm r},\alpha)}\begin{pmatrix}1 & -\ci \\ -\ci & -1\end{pmatrix}, 
    \label{eq:antiparallel-green-form}
\end{align}
from which 
\begin{align}
  \average{g}^2-\average{f}\average{\hconj{f}}&=0
  ,\\
  \check\Sigma^{\text{(a)}}(\epsilon,\bm{r}) &= \frac{\Gamma_{\text{n}}\average{g}}{\cpi\cos^2\delta_0}\begin{pmatrix}1 & -\ci \\ -\ci & -1\end{pmatrix}
    \label{eq:self-energy-green-function-a}
    ,
\end{align}
and 
\begin{equation}
{\tilde\Sigma}^{\text{(a)}}(\epsilon,\bm{r})=0,\quad E_{\Sigma}^{\text{(a)}}=0
  \label{eq:tilde-self-energy-zero}
\end{equation}
follow. The imaginary part of the second equation in \eqref{eq:tilde-self-energy-zero} yields $\Gamma^{\text{(a)}}(b)=0$, which reproduces in a simplified way the numerical results that $\Gamma$ is very small for the antiparallel vortex for $\delta_0\in\LeftClosedRightOpenInterval{0}{\delta_{\text{c}}}$. 
The vanishingly small $\Gamma$ has been expected in the Born limit. Our numerical results and a brief argument in this subsection show that $\Gamma$ is small (compared with $\Gamma_{\text{n}}$) for the antiparallel vortex with impurities with $\delta_0\in \LeftClosedRightOpenInterval{0}{\delta_{\text{c}}}$ as well as in the Born limit. 

We have confirmed that even when the anisotropic part (the partial wave with $\ell\ge 1$) of impurity scattering is taken into account, \eqref{eq:Sigma-diagonal}--\eqref{eq:Sigma-off-diagonal-2} hold owing to the rotational symmetry; the suppression of the impurity effect still takes place as long as $\Sigma_{\text{d}}$ is small compared with $\Delta_{\text{b}}$.

We should mention two drawbacks of the above argument. First, our numerical results show that $\Gamma$ in this case is indeed small but finite. The origin of this is not yet identified. Second, $E_{\Sigma}^{\text{(a)}}=0$ yields no renormalization of the excitation spectrum $\tilde{E}(b)$ of the bound states. Our numerical calculation, however, shows that $\tilde{E}(b)$ is renormalized by the presence of impurities. Thus, the improvement of the approximate analytical expression for the vortex bound state in the presence of impurities is a future problem.

\subsection{Impurity effect in parallel vortex for $\delta_0\in [0,\delta_{\text{c}}]$}
\label{subsection:small-delta0}
When we apply the argument in the previous subsection to the parallel vortex ($l=2$), it can be shown by the same approximation [that uses $\check{g}_{\text{cont}}\sim 0$ and \eqref{eq:perturbation-vortex-green-impure}] that $\tilde\Sigma^{\text{(p)}}$ is not zero, and therefore, in general, $E^{\text{(p)}}_{\Sigma}$ is not zero. The resultant $\Gamma$ is on the order of $\Gamma_{\text{n}}$, which is consistent with our numerical results.
The argument relying on a variant of the Kramer-Pesch approximation leads to a quantitative difference in $\Gamma$ between antiparallel and parallel vortices for $\delta_0( <\delta_{\text{c}})$.

The nonmonotonic $\delta_0$ dependence of $\Gamma$ can be obtained qualitatively within a non-self-consistent treatment.
We assume some explicit form of the pair potential \(\Delta\) and its characteristic length of variation \(\xi_1\) (let \(\Delta_{\text{b}}\) be the magnitude of \(\Delta\) in the bulk; assume that it is sufficiently large compared with \(\Gamma_{\text{n}}\)), ignore the induced component of the pair potential \(\Delta^{\text{(p)}}_{-}\), and retain the leading part in physical quantities with respect of \(b/\xi_1\) (these procedures are sometimes used for analyzing vortices: e.g., Refs.~\citen{Kato2000} and \citen{Melnikov2008}). We obtain \(u(s;b)=u(r)\) and
\begin{align}
  C^{\text{(p)}}(b) &= C_0 + \Order{(b/\xi_1)^2}
  ,\\
  E^{\text{(p)}}_\Delta(b) &= E_{\text{p}}b/\xi_1+\Order{(b/\xi_1)^3}, 
\end{align}
where \(C_0/\xi_1\) and \(E_{\text{p}}/\Delta_{\text{b}}\) are both on the order of unity. 
The averages of the Green's functions are then
\begin{align}
  \average{g(\epsilon,\bm{r},\alpha)}_{\alpha}
  &=
  \frac{-\ci\cpi \tilde{\xi}_1\exp[-u(r)]\uptheta(r-|b_\epsilon|)}{2\sqrt{r^2-b_\epsilon^2}}
  +\Order{(\epsilon/E_{\text{p}})^2}
  ,\\
  \average{f(\epsilon,\bm{r},\alpha)}_{\alpha}
  &=
\ci \average{g(\epsilon,\bm{r},\alpha)}_{\alpha}
\left( \frac{2b_\epsilon^2-r^2}{r^2}\right)
  \ce^{+2\ci\phi}
  +\Order{(\epsilon/E_{\text{p}})^2}
  ,\\
  -\average{\hconj{f}(\epsilon,\bm{r},\alpha)}_{\alpha}
  &=
\ci \average{g(\epsilon,\bm{r},\alpha)}_{\alpha}
\left( \frac{2b_\epsilon^2-r^2}{r^2}\right)
  \ce^{-2\ci\phi}
  +\Order{(\epsilon/E_{\text{p}})^2}
  ,
\end{align}
where $\tilde\xi_1 = v_{\text{F}}\xi_1/(C_0E_{\text{p}})$ and $b_\epsilon=\xi_1\epsilon/E_{\text{p}}$. To derive the above, we use the relationship \((z+\ci\eta)^{-1}=-\ci\cpi\delta(z) + \operatorname{P}z^{-1}\) and ignore the latter principal part \(\operatorname{P}z^{-1}\). This also implies that we disregard the renormalization factor and let \(Z=1\) in \eqref{eq:gamma-from-esigma-and-z}.
The denominator of \eqref{eq:self-energy-green-function} becomes
\begin{align}
  \cos^2\delta_0-\frac{\sin^2\delta_0}{\cpi^2}\left(\average{g}^2-\average{f}\average{\hconj{f}}\right)
  &=\cos^2\delta_0+\sin^2\delta_0\frac{ \tilde{\xi}_1^2 b_\epsilon^2}{r^4}
  \exp[-2u(r)]\uptheta(r-b_\epsilon)
  +\Order{(\epsilon/E_{\text{p}})^2}.
\end{align}

Using these quantities, we can obtain the self-energy $\tilde\Sigma^{\text{(p)}}$.  
We integrate $\tilde\Sigma^{\text{(p)}}$ along the trajectory with the impact parameter $b_\epsilon$ as in \eqref{eq:esigma-from-tilde-sigma-by-integration}, and then obtain an approximate $\Gamma^{\text{(p)}}$ (notice that \(\exp[\ci(\phi-\alpha)]=(s+\ci b)/r\)) for $\bm{r}$ on the trajectory satisfying $b=b_{\epsilon}$,
\begin{widetext}
\begin{align}
  \Gamma(\epsilon,b=b_\epsilon)
  &=
  2\Gamma_{\text{n}}\frac{\tilde{\xi}_1}{C_0}
\int_0^\infty\dd s\frac{ s b_\epsilon^2 \exp[-2u(\sqrt{s^2+b_\epsilon^2})]
}
{\cos^2\delta_0 (s^2+b_\epsilon^2)^2 +\sin^2\delta_0 (\tilde{\xi}_1^2 b_\epsilon^2)\exp[-2u(\sqrt{s^2+b_\epsilon^2})]}.
  \label{eq: gamma-integral}
\end{align}
\end{widetext}
This integral is not performed analytically; as a rough evaluation, we can approximate $\exp[-2u(\sqrt{s^2+b_\epsilon^2})]$ as $\uptheta(\xi_{\text{c}}-|s|)$ with a cutoff $\xi_{\text{c}}$ on the order of $\xi_1$. 
The integral in \eqref{eq: gamma-integral} then reduces to
\begin{align}
\Gamma(\epsilon,b=b_\epsilon)
&\sim
2\Gamma_{\text{n}}\frac{\tilde{\xi}_1}{C_0}
\int_0^{\xi_{\text{c}}}\dd s\frac{ s b_\epsilon^2}
{\cos^2\delta_0 (s^2+b_\epsilon^2)^2 +\sin^2\delta_0 (\tilde{\xi}_1^2 b_\epsilon^2)} \label{eq: gamma-integral-reduction-1}\\
&=
\frac{2\Gamma_{\text{n}} b_\epsilon}{C_0\sin2\delta_0}
\left\{\arctan\left[\frac{\xi_{\text{c}}^2+b_\epsilon^2}{\tilde{\xi}_1b_\epsilon\tan\delta_0}\right]-\arctan\left[\frac{b_\epsilon}{\tilde{\xi}_1\tan\delta_0}\right]\right\}.
\label{eq: gamma-integral-reduction}
\end{align}
As a reference, the expression for the \textit{s}-wave vortex is given in Appendix \ref{appendix:s-wave-non-self-consistent-gamma}
.
When $\delta_0=0$, \eqref{eq: gamma-integral-reduction} becomes
\begin{align}
  \Gamma(\epsilon,b=b_\epsilon) & \underset{\delta_0\to 0}{=}
  2\Gamma_{\text{n}}\frac{\tilde\xi_1}{C_0}\int_0^{\xi_{\text{c}}}\dd s\frac{s b_\epsilon^2}{(s^2+b_\epsilon^2)^2}
  \\
  &=
  \frac{\Gamma_{\text{n}}\tilde\xi_1\xi_{\text{c}}^2}{C_0(\xi_{\text{c}}^2+b_\epsilon^2)}.
  \label{eq: gamma-integral-reduction-limits-first-delta0}
\end{align}
Note that $\tilde{\xi}_1/C_0 \sim \Order{1}$ and the Kramer-Pesch approximation written down in Sec. 2.4 is valid only for $b_\epsilon \ll \xi_{\text{c}}$ (which corresponds to $\epsilon \ll \Delta_{\text{b}}$). We then obtain 
\begin{equation}
\Gamma(\epsilon,b=b_\epsilon)\Big|_{\delta_0=0}\sim \Gamma_{\text{n}}. 
\label{eq: Gamma-p-estimation}
\end{equation}
When \(\delta_0\) is nonzero, on the other hand, we obtain
\begin{align}
  \eqref{eq: gamma-integral-reduction}
  &\underset{b_\epsilon\to 0}{=} 0, &\text{for all \(\delta_0 > 0\)},
  \label{eq: gamma-integral-reduction-limits-first-bepsilon}
\end{align}
and this seems to conflict with our numerical results. 
This problem can be resolved by considering \(\Gamma\) of finite \(b\).
As a remedy, we assume that 
\begin{align}
&\Gamma(\epsilon,b=b_\epsilon)\text{ in self-consistent calculation}\nonumber\\
&\quad\sim \begin{cases}
\Gamma(\epsilon,b_\epsilon)\text{ in non-self-consistent calculation},&
\epsilon\geq\epsilon^{*}
\\
\Gamma(\epsilon^{*},b_{\epsilon^{*}})\text{ in non-self-consistent calculation},&
\epsilon\leq\epsilon^{*}
\end{cases}\label{eq: prescription}
\end{align}
where \(\epsilon^{*}>0\) is defined as the minimum \(\epsilon\) that satisfies \(\Gamma(\epsilon,b=b_{\epsilon}) = \epsilon\).
Figure \ref{fig:parallel-gamma-too-simple} shows \(\epsilon^{*} = \Gamma(\epsilon^{*},b_{\epsilon^{*}})\) calculated from Eq. \eqref{eq: gamma-integral-reduction} with \(C_0=\xi_{\text{c}}=\tilde\xi_1=\xi_0\), \(E_{\text{p}}=\Delta_0\), and \(\Gamma_{\text{n}}=0.2\Delta_0\). 
This assumption qualitatively well reproduces its dependence on \(\delta_0\) although there are two points to be improved. (i) $\Gamma(\epsilon^{*},b=b_{\epsilon^{*}})$ becomes zero while a finite $\Gamma$ is numerically obtained. (ii) $\Gamma(\epsilon^{*},b=b_{\epsilon^{*}})$ for $\delta_0=0$ is reduced, in comparison with $\Gamma_{\text{n}}$. 

The suppression of \(\Gamma\) by the finite \(\delta_0\) can be explained with \eqref{eq:self-energy-green-function}; the increase in \(\delta_0\) increases the denominator of the self-energy in \eqref{eq:self-energy-green-function} and therefore suppresses \(\Sigma\) (or \(\tilde\Sigma\)) within the vortex core.

\begin{figure}
\centering
\includegraphics[width=0.4\textwidth]{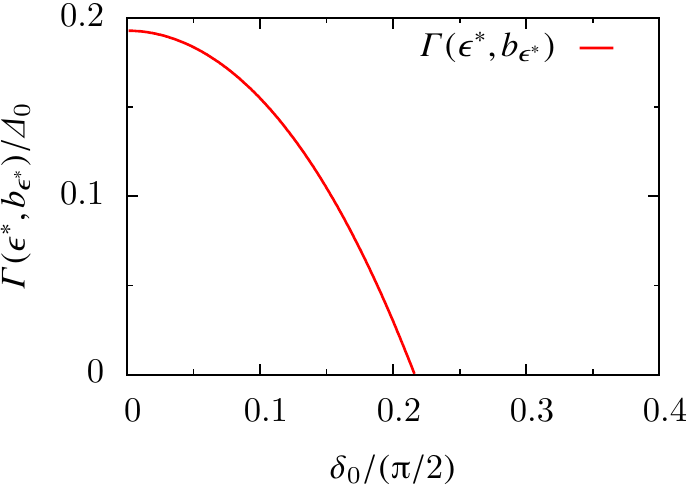}
\caption{(Color online) $\Gamma(\epsilon^{*},b_{\epsilon^{*}}) = \epsilon^{*}$ from \eqref{eq: gamma-integral-reduction} with \(C_0=\xi_{\text{c}}=\tilde\xi_1=\xi_0\), \(E_{\text{p}}=\Delta_0\), and \(\Gamma_{\text{n}}=0.2\Delta_0\). }
\label{fig:parallel-gamma-too-simple}
\end{figure}

\section{Conclusion}

In this work, we study the phase-shift dependence of impurity scattering on the vortices of two-dimensional chiral $p$-wave superconductors. We found the critical value \(\delta_{\text{c}}\), above which the spectrum is heavily broadened for both parallel and antiparallel vortices. We attribute this to the mixing between the vortex-ABS and the impurity bands in the bulk. For \(\delta_0 < \delta_{\text{c}}\), we confirm the robustness of antiparallel vortices. We also found that a finite \(\delta_0\) suppresses the impurity effect within the parallel vortex core. Some of the above findings can be explained qualitatively by the non-self-consistent Kramer-Pesch approximation.

\begin{acknowledgments}
  We thank J.\ A.\ Sauls for helpful discussions and E.\ Arahata for preliminary numerical calculations.
  This work was supported by JSPS KAKENHI Grant Number 23244070 and 15K05160.
\end{acknowledgments}

\appendix

\section{Derivation of \eqref{eq:critical-phase-shift}}
\label{appendix:critical-phase-shift}
Equation \eqref{eq:critical-phase-shift} can be derived from the bulk solution of the diagonal part of the quasiclassical Green's function $g$,
\begin{align}
  g(z) = \frac{-\cpi\tilde z}{\sqrt{-\tilde z^2+\abs{\Delta}^2}}
  .
\end{align}
From \eqref{eq:self-energy-green-function}, the self-energy in the bulk is
\begin{align}
  \check\Sigma(z) = \frac{\Gamma_{\text{n}}\average{g(z)}\check\tau_3/\cpi}{\cos^2\delta_0-\sin^2\delta_0\average{g(z)}^2/\cpi^2}
  ,
\end{align}
and because $\average{g}=g$, the diagonal part of the quasiclassical self-energy $\Sigma_{\text{d}}(\ci\omega_n)$ for $\ci\omega_n\to0+\ci\eta$, the imaginary part of which gives the DOS at the Fermi level, satisfies the sextic equation
\begin{align}
  \Sigma_{\text{d}}^2\left\{\Sigma_{\text{d}}^4+\left(\Gamma_{\text{n}}^2-2\abs{\Delta}^2\cos^2\delta_0\right)\Sigma_{\text{d}}^2
  +\left(\abs{\Delta}^4\cos^4\delta_0-\Gamma_{\text{n}}^2\abs{\Delta}^2\right)\right\}
  = 0
  .
\end{align}
Thus, the behavior of the root of the quadratic equation
\begin{align}
  x^2+(\Gamma_{\text{n}}^2-2\abs{\Delta}^2\cos^2\delta_0)x+(\abs{\Delta}^4\cos^4\delta_0-\Gamma_{\text{n}}^2\abs{\Delta}^2)=0
  \label{eq:quadratic-equation}
\end{align}
tells us the existence of the DOS on the Fermi surface.
We know that $\Re\Sigma_{\text{d}}=0$ because of the particle-hole symmetry. In addition, we also know that $\Im\Sigma_{\text{d}}=0$ when $\Gamma_{\text{n}}=0$ and $\delta_0=0$, and $\Im\Sigma_{\text{d}}\ne 0$ when $\delta_0=\cpi/2$ with finite $\Gamma_{\text{n}}$\cite{Maki1999, Maki2000}. The physical solution should continuously vary from these limits when $\Gamma_{\text{n}}$ or $\delta_0$ or both are continuously changing.
Consequently, the condition under which \eqref{eq:quadratic-equation} has negative real solutions is
\begin{align}
  \cos\delta_0 &< \sqrt{\Gamma_{\text{n}}/\abs{\Delta}} &\text{or}& & \sqrt{2}\cos\delta_0 &< \Gamma_{\text{n}}/\abs{\Delta}
  ,
\end{align}
and in both cases, \eqref{eq:quadratic-equation} has only one negative solution.
For temperatures near $T_{\text{c}}$, $\Gamma_{\text{n}} > \abs{\Delta}$ and both equations are satisfied; there is an impurity band.
For $\Gamma_{\text{n}} \le \abs{\Delta}$ (which means sufficiently low temperature and low impurity scattering in the normal state),
\begin{align}
  \delta_0 > \arccos\sqrt{\Gamma_{\text{n}}/\abs{\Delta}}
\end{align}
is a necessary condition for nonzero $\Sigma_{\text{d}}$.
Assuming that $\Delta$ is a monotonically decreasing function of $\delta_0$ for given $\Gamma_\text{n}$ (we check this statement numerically for the parameters used in this paper, as in Fig.~\ref{fig:pair-potential-vs-phase-shift}), $\delta_{\text{c}}$ defined as \eqref{eq:critical-phase-shift} is unique and $\Sigma_{\text{d}}$ has only one nonzero solution reachable from the limits when $\delta_0$ is greater than $\delta_{\text{c}}$ and $\Gamma_{\text{n}} \le \abs{\Delta}$. Therefore, at low temperatures, it is a critical point of the phase-shift, above which there exists a finite DOS on the Fermi surface.

%
%
%
\section{Spectrum at the vortex core}
\label{appendix:supplementary}

In this appendix, we present the spectrum of angular resolved LDOS at the vortex core for some parameters (Figs.~\ref{fig:appendix:angular-resolved-center-l0t01g02}--\ref{fig:appendix:angular-resolved-center-l2t01g04}).
In these figures, the dotted lines represent the fitted Lorentzian functions in the same way as in Fig.~\ref{fig:bulk-core-t03g03m}. We can see the following properties: 
\begin{enumerate}[(i)]
\item The nonmonotonic \(\delta_0\)-dependence of \(\Gamma_{\text{n}}\) of the parallel vortex is not negligible for a clean SC (Fig.~\ref{fig:appendix:angular-resolved-center-l2t01g02}). 
\item The Lorentzian fitting of the spectrum does not work well for strong scatterers (large \(\delta_0\) for antiparallel case, and large \(\delta_0\) or large \(\Gamma_{\text{n}}\) for parallel case), where the structure of the spectrum is nearly lost.
\end{enumerate}
\begin{figure*}
\centering
\includegraphics[width=0.32\textwidth]{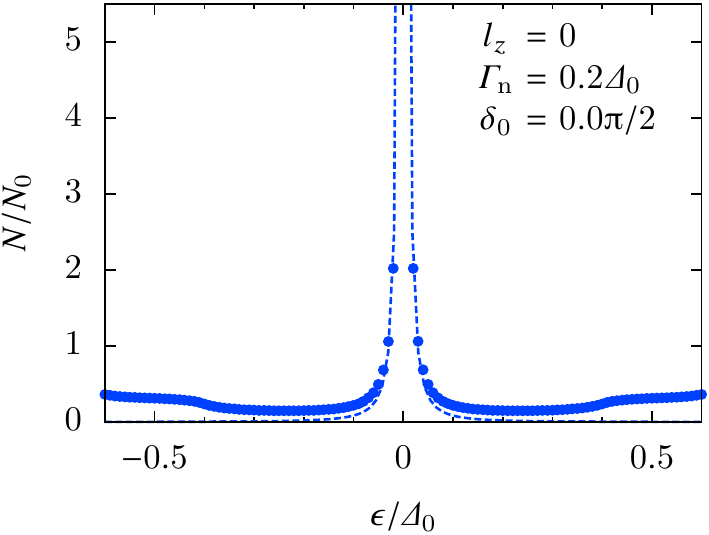}
\includegraphics[width=0.32\textwidth]{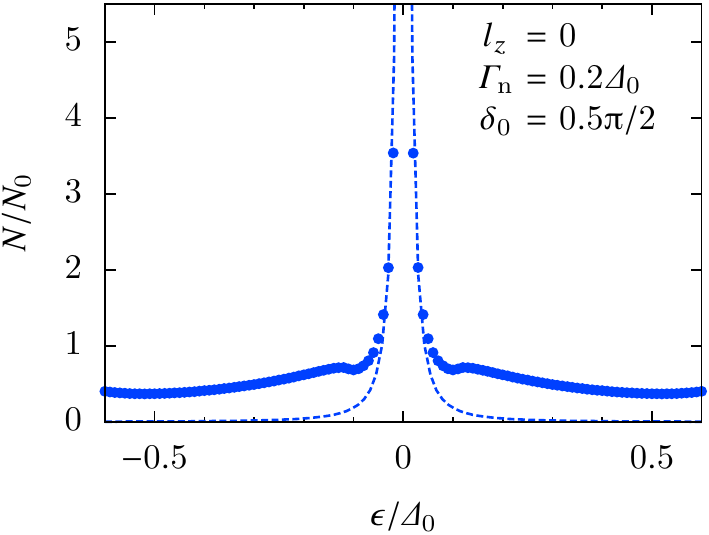}
\includegraphics[width=0.32\textwidth]{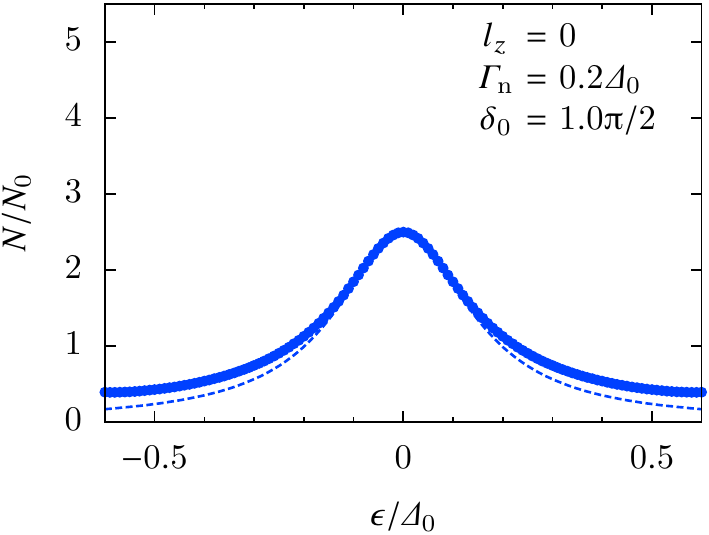}
\caption{(Color online) Angular resolved LDOS at vortex core for \(l_z=0\) (antiparallel), \(T=0.1T_{\text{c0}}\), \(\Gamma_{\text{n}}=0.2\Delta_0\). Left: \(\delta_0=0.0\cpi/2\) (Born limit); center: \(\delta_0=0.5\cpi/2\); right: \(\delta_0=1.0\cpi/2\) (unitary limit).}
\label{fig:appendix:angular-resolved-center-l0t01g02}
\end{figure*}
\begin{figure*}
\centering
\includegraphics[width=0.32\textwidth]{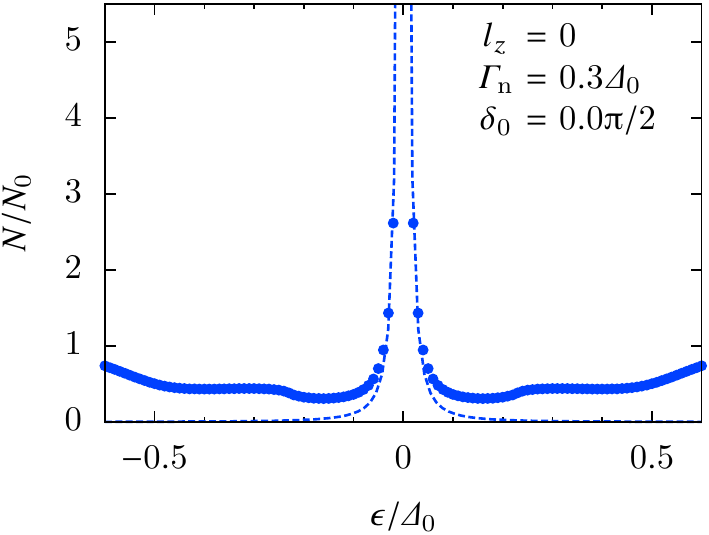}
\includegraphics[width=0.32\textwidth]{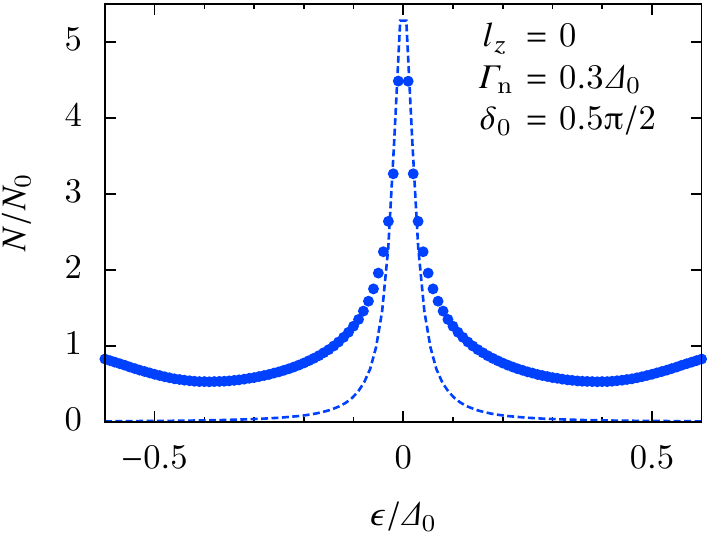}
\includegraphics[width=0.32\textwidth]{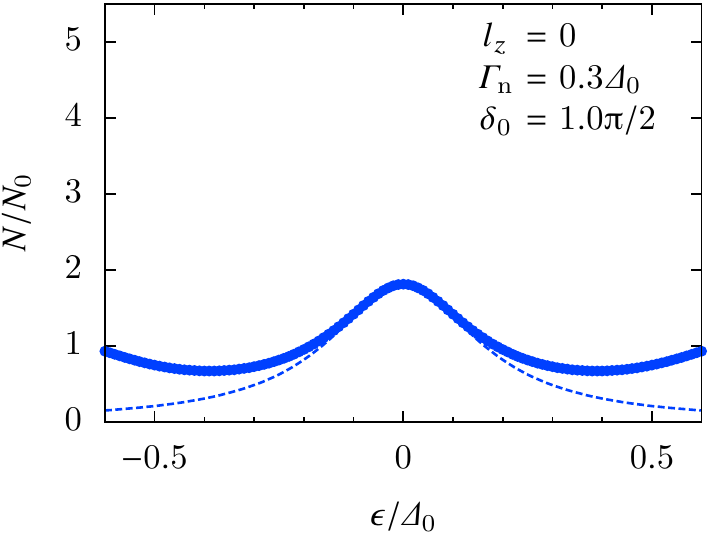}
\caption{(Color online) Angular resolved LDOS at vortex core for \(l_z=0\) (antiparallel), \(T=0.1T_{\text{c0}}\), \(\Gamma_{\text{n}}=0.3\Delta_0\). Left: \(\delta_0=0.0\cpi/2\) (Born limit); center: \(\delta_0=0.5\cpi/2\); right: \(\delta_0=1.0\cpi/2\) (unitary limit).}
\label{fig:appendix:angular-resolved-center-l0t01g03}
\end{figure*}
\begin{figure*}
\centering
\includegraphics[width=0.32\textwidth]{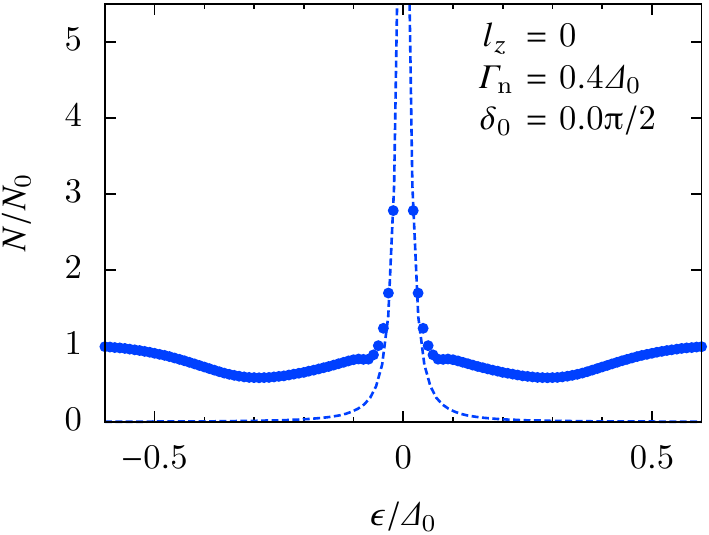}
\includegraphics[width=0.32\textwidth]{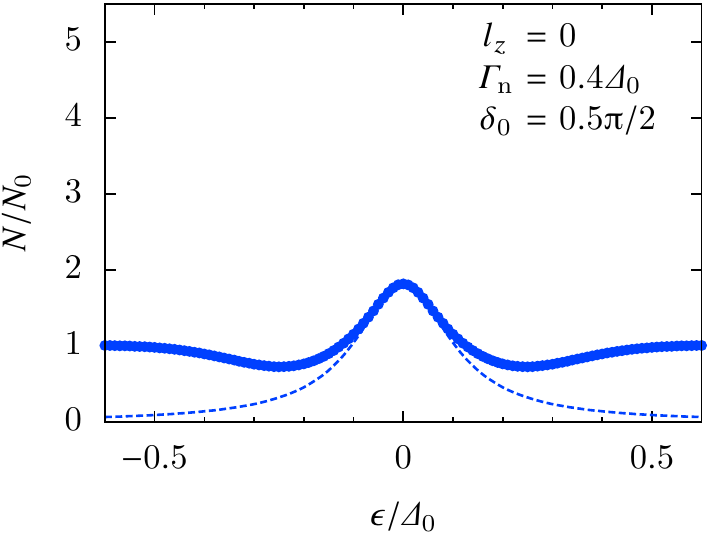}
\includegraphics[width=0.32\textwidth]{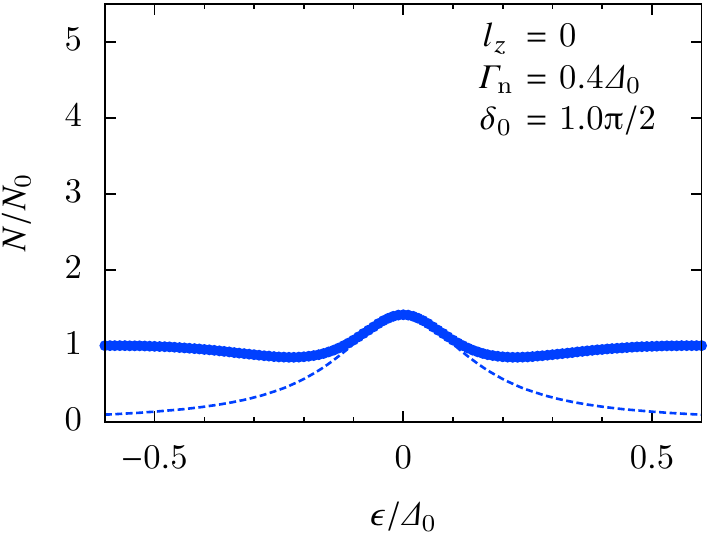}
\caption{(Color online) Angular resolved LDOS at vortex core for \(l_z=0\) (antiparallel), \(T=0.1T_{\text{c0}}\), \(\Gamma_{\text{n}}=0.4\Delta_0\). Left: \(\delta_0=0.0\cpi/2\) (Born limit); center: \(\delta_0=0.5\cpi/2\); right: \(\delta_0=1.0\cpi/2\) (unitary limit).}
\label{fig:appendix:angular-resolved-center-l0t01g04}
\end{figure*}
\begin{figure*}
\centering
\includegraphics[width=0.32\textwidth]{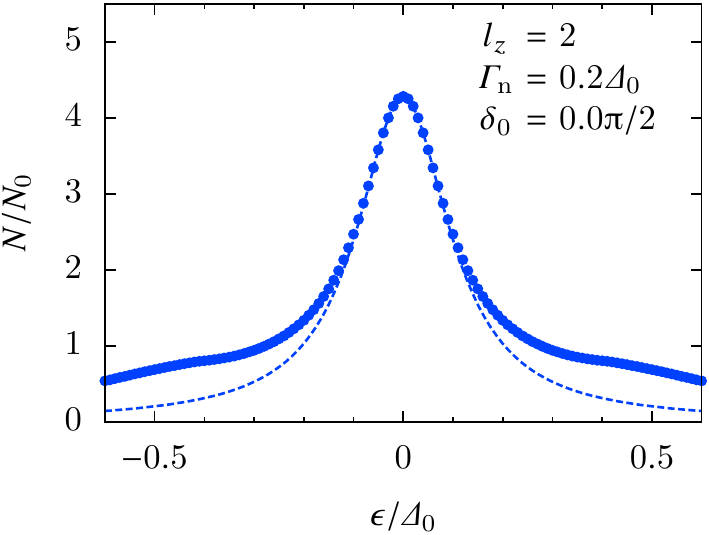}
\includegraphics[width=0.32\textwidth]{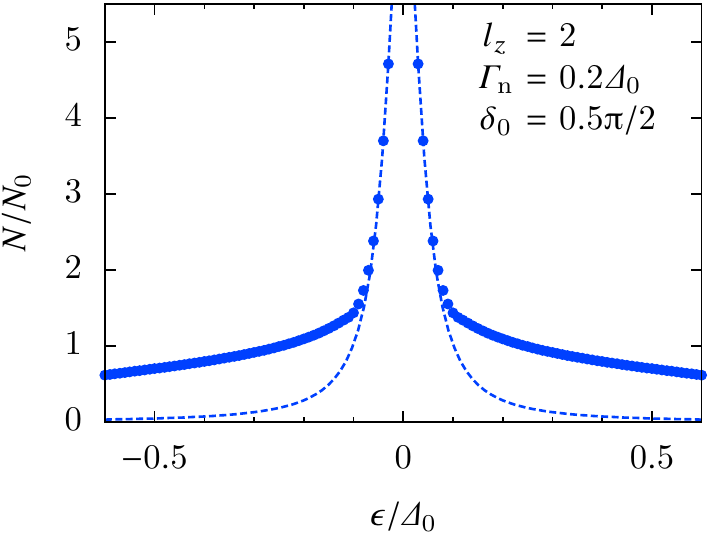}
\includegraphics[width=0.32\textwidth]{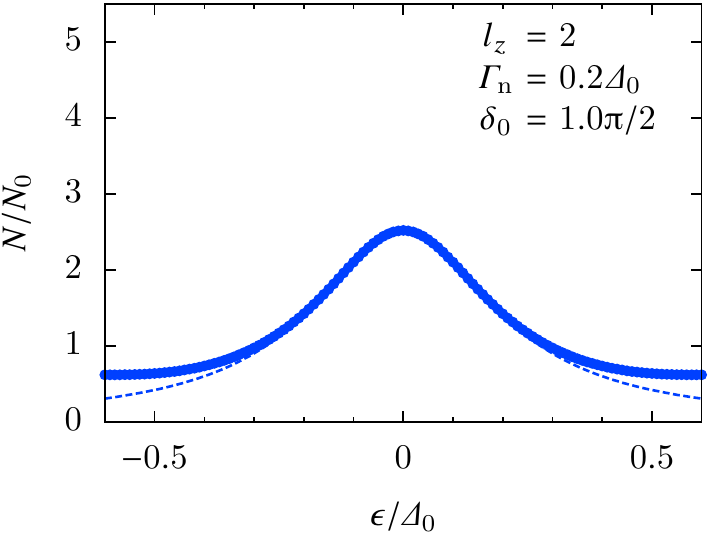}
\caption{(Color online) Angular resolved LDOS at vortex core for \(l_z=2\) (parallel), \(T=0.1T_{\text{c0}}\), \(\Gamma_{\text{n}}=0.2\Delta_0\). Left: \(\delta_0=0.0\cpi/2\) (Born limit); center: \(\delta_0=0.5\cpi/2\); right: \(\delta_0=1.0\cpi/2\) (unitary limit).}
\label{fig:appendix:angular-resolved-center-l2t01g02}
\end{figure*}
\begin{figure*}
\centering
\includegraphics[width=0.32\textwidth]{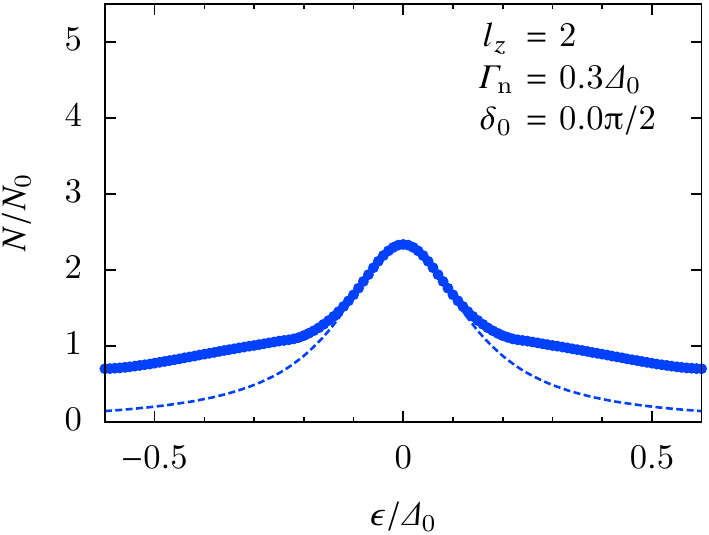}
\includegraphics[width=0.32\textwidth]{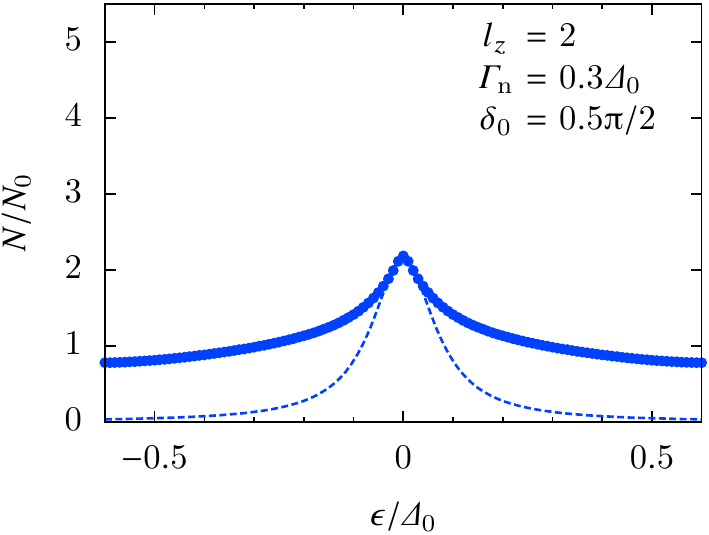}
\includegraphics[width=0.32\textwidth]{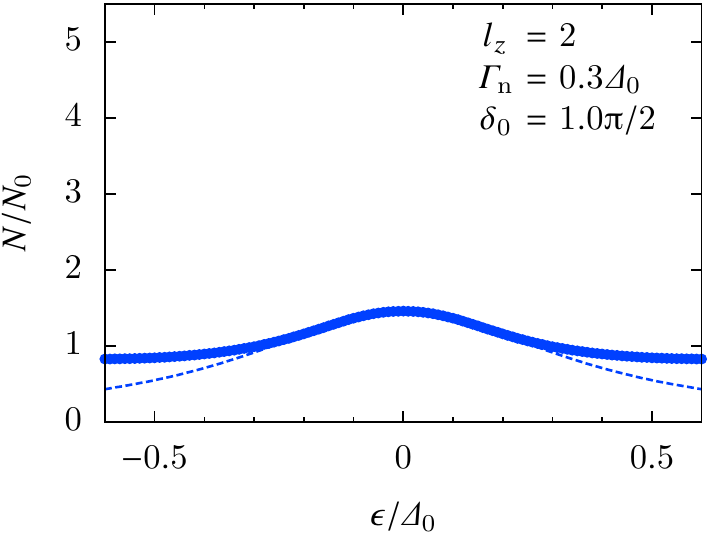}
\caption{(Color online) Angular resolved LDOS at vortex core for \(l_z=2\) (parallel), \(T=0.1T_{\text{c0}}\), \(\Gamma_{\text{n}}=0.3\Delta_0\). Left: \(\delta_0=0.0\cpi/2\) (Born limit); center: \(\delta_0=0.5\cpi/2\); right: \(\delta_0=1.0\cpi/2\) (unitary limit).}
\label{fig:appendix:angular-resolved-center-l2t01g03}
\end{figure*}
\begin{figure*}
\centering
\includegraphics[width=0.32\textwidth]{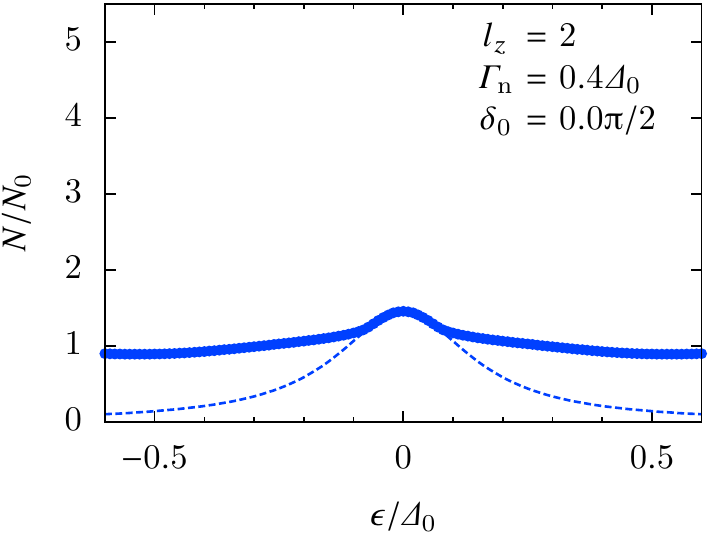}
\includegraphics[width=0.32\textwidth]{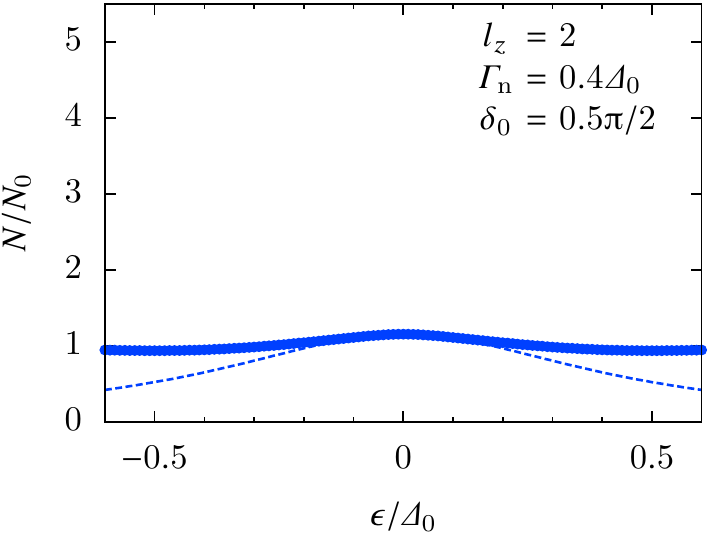}
\includegraphics[width=0.32\textwidth]{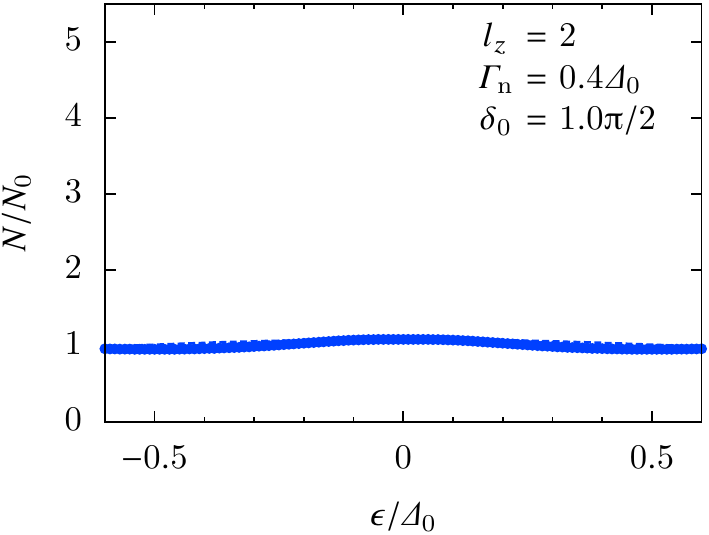}
\caption{(Color online) Angular resolved LDOS at vortex core for \(l_z=2\) (parallel), \(T=0.1T_{\text{c0}}\), \(\Gamma_{\text{n}}=0.4\Delta_0\). Left: \(\delta_0=0.0\cpi/2\) (Born limit); center: \(\delta_0=0.5\cpi/2\); right: \(\delta_0=1.0\cpi/2\) (unitary limit).}
\label{fig:appendix:angular-resolved-center-l2t01g04}
\end{figure*}

\section{Phase-shift dependence of $s$-wave vortices}
\label{appendix:s-wave-non-self-consistent-gamma}
In the \(s\)-wave case, the same procedure yields
\begin{align}
  \average{g(\epsilon,\bm{r},\alpha)}_\alpha
  &=
  \frac{-\ci\cpi\tilde\xi_1\exp[-u(r)]\uptheta(r-\abs{b_\epsilon})}{2\sqrt{r^2-b_\epsilon^2}}
  +\Order{(\epsilon/E_{\text{p}})^2}
  ,\\
  \average{f(\epsilon,\bm{r},\alpha)}_\alpha
  &=
  -\average{g(\epsilon,\bm{r},\alpha)}_\alpha\frac{b_\epsilon}{r}\ce^{+\ci\phi}
  +\Order{(\epsilon/E_{\text{p}})^2}
  ,\\
  -\average{\hconj{f}(\epsilon,\bm{r},\alpha)}_\alpha
  &=
  +\average{g(\epsilon,\bm{r},\alpha)}_\alpha\frac{b_\epsilon}{r}\ce^{-\ci\phi}
  +\Order{(\epsilon/E_{\text{p}})^2}
  ,
\end{align}
and
\begin{align}
  \cos^2\delta_0-\frac{\sin^2\delta_0}{\cpi^2}\left(\average{g}^2-\average{f}\average{\hconj{f}}\right)
  &=
  \cos^2\delta_0+\sin^2\delta_0\frac{\tilde\xi_1^2}{4r^2}\exp[-2u(r)]\uptheta(r-b_\epsilon)
  +\Order{(\epsilon/E_{\text{p}})^2}
  .
\end{align}
The scattering rate is then
\begin{align}
\Gamma(\epsilon,b=b_{\epsilon})
&\sim
-\Im\int_0^{\xi_{\text{c}}}\dd s\frac{\tilde\Sigma}{C_0}
\\
&\sim
\frac{2\Gamma_{\text{n}}\tilde\xi_1}{C_0}\int_0^{\xi_{\text{c}}}\dd s\frac{s}{4(s^2+b_\epsilon^2)\cos^2\delta_0+\tilde\xi_1^2\sin^2\delta}
\\
&=
\Gamma_{\text{n}}\frac{\tilde\xi_1}{4C_0}\frac{1}{\cos^2\delta_0}\ln\frac{4(\xi_{\text{c}}^2+b_\epsilon^2)\cos^2\delta_0+\tilde\xi_1^2\sin^2\delta_0}{4b_\epsilon^2\cos^2\delta_0+\tilde\xi_1^2\sin^2\delta_0}.
\end{align}
In the limit \(\delta_0\to 0\), it becomes\cite{Kopnin1995}
\begin{align}
\Gamma&=
\frac{\Gamma_{\text{n}}\tilde\xi_1}{4C_0}\ln\frac{\xi_{\text{c}}^2+b_\epsilon^2}{b_\epsilon^2}
\nonumber\\
&\simeq
-\frac{\Gamma_{\text{n}}\tilde\xi_1}{2C_0}\ln\frac{\abs{b_\epsilon}}{\xi_{\text{c}}}
& \text{for $\abs{b_{\epsilon}}\ll \xi_{\text{c}}\sim\xi_0$} 
,
\end{align}
and in the limit \(b_\epsilon\to 0\),
\begin{align}
\Gamma&=
\frac{\Gamma_{\text{n}}\tilde\xi_1}{4C_0}\frac{1}{\cos^2\delta_0}\ln\frac{4\xi_{\text{c}}^2\cos^2\delta_0+\tilde\xi_1^2\sin^2\delta_0}{\tilde\xi_1^2\sin^2\delta_0}
\nonumber\\
&\simeq
-\frac{\Gamma_{\text{n}}\tilde\xi_1}{2C_0\cos^2\delta_0}\ln\delta_0
& \text{for $\delta_0\to 0$}
\label{eq:suppression-by-phase-shift-in-swave}
.
\end{align}
Equation \eqref{eq:suppression-by-phase-shift-in-swave} shows that in the $s$-wave superconductors with the non-self-consistent treatment of impurity self-energy, a finite phase-shift suppresses the logarithmic divergence at the vortex core, as in a previous study\cite{Koulakov1999}.


\begin{thebibliography}{99}
\bibitem{Caroli1964} C.\ Caroli, P.\ G.\ de~Gennes, and J.\ Matricon, Phys.\ Lett.\ \textbf{9}, 307 (1964).
\bibitem{Stone1996} M.\ Stone, Phys.\ Rev.\ B \textbf{54}, 13222 (1996).
%
\bibitem{Read2000} N.\ Read and D.\ Green, Phys.\ Rev.\ B \textbf{61}, 10267 (2000).
\bibitem{Ivanov2001} D.\ A.\ Ivanov, Phys.\ Rev.\ Lett.\ \textbf{86}, 268 (2001).
\bibitem{Nayak2008} C.\ Nayak, S.\ H.\ Simon, A.\ Stern, M.\ Freedman, and S.\ D.\ Sarma, Rev.\ Mod.\ Phys.\ \textbf{80}, 1083 (2008).
%
\bibitem{Vorontsov2003} A.\ B.\ Vorontsov and J.\ A.\ Sauls, Phys.\ Rev.\ B \textbf{68}, 064508 (2003).
\bibitem{Levitin2013} L.\ V.\ Levitin, R.\ G.\ Bennett, A.\ Casey, B.\ Cowan, J.\ Saunders, D.\ Drung, Th.\ Schurig, and J.\ M.\ Parpia, Science \textbf{340}, 841 (2013).
\bibitem{Maeno1994} Y.\ Maeno, H.\ Hashimoto, K.\ Yoshida, S.\ Nishizaki, T.\ Fujita, J.\ G.\ Bednorz, and F.\ Lichtenberg, Nature (London) \textbf{372}, 532 (1994).
\bibitem{Mackenzie2003} A.\ P.\ Mackenzie and Y.\ Maeno, Rev.\ Mod.\ Phys.\ \textbf{75}, 657 (2003).
\bibitem{Sigrist2005} M.\ Sigrist, Prog.\ Theor.\ Phys.\ Suppl.\ \textbf{160}, 1 (2005).
\bibitem{Maeno2012} Y.\ Maeno, S.\ Kittaka, T.\ Nomura, S.\ Yonezawa, and K.\ Ishida, J.\ Phys.\ Soc.\ Jpn.\ \textbf{81}, 011009 (2012).
\bibitem{Salmelin1989} R.\ H.\ Salmelin, M.\ M.\ Salomaa, and V.\ P.\ Mineev, Phys.\ Rev.\ Lett.\ \textbf{63}, 868 (1989).
\bibitem{Salmelin1990} R.\ H.\ Salmelin and M.\ M.\ Salomaa, Phys.\ Rev.\ Lett.\ \textbf{41}, 4142 (1990).
\bibitem{Ikegami2013} H.\ Ikegami, Y.\ Tsutsumi, and K.\ Kono, Science \textbf{341}, 59 (2013).
%
\bibitem{Tanuma2009} Y.\ Tanuma, N.\ Hayashi, Y.\ Tanaka, and A.\ A.\ Golubov, Phys.\ Rev.\ Lett.\ \textbf{102}, 117003 (2009).
\bibitem{Tokuyasu1990} T.\ A.\ Tokuyasu, D.\ W.\ Hess, and J.\ A.\ Sauls, Phys.\ Rev.\ B \textbf{41}, 8891 (1990).
\bibitem{Matsumoto2000} M.\ Matsumoto and M.\ Sigrist, Physica B \textbf{281-282}, 973 (2000).
\bibitem{Matsumoto2001} M.\ Matsumoto and R.\ Heeb, Phys. Rev.\ B \textbf{65}, 014504 (2001).
\bibitem{Kato2000} Y.\ Kato, J.\ Phys.\ Soc.\ Jpn.\ \textbf{69}, 3378 (2000).
\bibitem{Kato2002} Y.\ Kato and N.\ Hayashi, J.\ Phys.\ Soc.\ Jpn.\ \textbf{71}, 1721 (2002).
\bibitem{Hayashi2005} N.\ Hayashi, Y.\ Kato, and M.\ Sigrist, J.\ Low\ Temp.\ Phys.\ \textbf{139}, 79 (2005).

\bibitem{Yokoyama2009} T.\ Yokoyama, C.\ Ichioka, Y.\ Tanaka, and M.\ Sigrist, Phys.\ Rev.\ B \textbf{78}, 012508 (2008).
\bibitem{Tanaka2012} Y.\ Tanaka, M.\ Sato, and N.\ Nagaosa, J.\ Phys.\ Soc.\ Jpn.\ \textbf{81}, 011013 (2012).
\bibitem{Volovik1999} G.\ E.\ Volovik, Pis'ma v ZhETF \textbf{70}, 601 (1999) [JETP Lett.\ \textbf{70}, 609 (1999)].
\bibitem{Hirschfeld1986} P.\ Hirschfeld, D.\ Vollhardt, and P.\ W\"olfle, Solid State Commun. \textbf{59}, 111 (1986).
\bibitem{SchmittRink1986} S.\ Schmitt-Rink, K.\ Miyake, and C.\ M.\ Varma, Phys.\ Rev.\ Lett.\ \textbf{57}, 2527 (1986).
\bibitem{Suzuki2002} M.\ Suzuki, M.\ A.\ Tanatar, N.\ Kikugawa, Z.\ Q.\ Mao, Y.\ Maeno, and T.\ Ishiguro, Phys.\ Rev.\ Lett.\ \textbf{88}, 227004 (2002).
\bibitem{Kramer1974} L.\ Kramer and W.\ Pesch, Z.\ Phys.\ \textbf{269}, 59 (1974).
\bibitem{Hayashi2013-pwave} N.\ Hayashi, N.\ Kurosawa, E.\ Arahata, Y.\ Kato, Y.\ Tanuma, Y.\ Tanaka, and A.\ A.\ Golubov, Physica C \textbf{494}, 131 (2013).
\bibitem{Sauls2009} J.\ A.\ Sauls and M.\ Eschrig, New J.\ Phys.\ \textbf{11}, 075008 (2009).

%
\bibitem{Kurosawa2014} N.\ Kurosawa, N.\ Hayashi, E.\ Arahata and Y.\ Kato, J.\ Low\ Temp.\ Phys.\ \textbf{175}, 365 (2014).
%
\bibitem{Eilenberger1968} G.\ Eilenberger, Z.\ Phys.\ \textbf{214}, 195 (1968).
\bibitem{Larkin1969} A.\ I.\ Larkin and Yu.\ N.\ Ovchinnikov, Zh.\ Eksp.\ Teor.\ Fiz.\ \textbf{55}, 2262 (1968) [Sov.\ Phys.\ JETP \textbf{28}, 1200 (1969)].
\bibitem{Serene1983} J.\ W.\ Serene and D.\ Rainer, Phys.\ Rep.\ \textbf{101}, 221 (1983).
%
\bibitem{Thuneberg1984} E.\ V.\ Thuneberg, J.\ Kurkij\"arvi, and D. Rainer, Phys.\ Rev.\ B \textbf{29}, 3913 (1984).
%
\bibitem{Larkin1965} A.\ I.\ Larkin, Zh.\ Eksp.\ Teor.\ Fiz.\ Pis'ma Red.\ \textbf{2}, 205 (1965) [JETP Lett.\ \textbf{2}, 130 (1965)].
\bibitem{Mineev} V.\ P.\ Mineev and K.\ V.\ Samokhin, \textit{Introduction to Unconventional Superconductivity} (Gordon and Breach Science Publishers, New York, 1999).
%
\bibitem{Maki1999} K.\ Maki and E.\ Puchkaryov, Europhys.\ Lett.\ \textbf{45}, 263 (1999).
\bibitem{Maki2000} K.\ Maki and E.\ Puchkaryov, Europhys.\ Lett.\ \textbf{50}, 533 (2000).






\bibitem{Nagato1993} Y.\ Nagato, K.\ Nagai, and J.\ Hara, J.\ Low\ Temp.\ Phys.\ \textbf{93}, 33 (1993).
\bibitem{Schopohl1995} N.\ Schopohl and K.\ Maki, Phys.\ Rev.\ B \textbf{52}, 490 (1995).
\bibitem{Schopohl1998} N.\ Schopohl, arXiv:cond-mat/9804064.

  
\bibitem{Shampine1986} L.\ F.\ Shampine, Math.\ Comp.\ \textbf{46}, 134 (1986).
\bibitem{Hayashi2013-swave} N.\ Hayashi, Y.\ Higashi, N.\ Nakai, and Y.\ Suematsu, Physica C \textbf{484}, 69 (2013).
\bibitem{Eyert1996} V.\ Eyert, J.\ Comput.\ Phys.\ \textbf{124}, 271 (1996).


\bibitem{Sauls2013} J.\ A.\ Sauls, private communication.


\bibitem{Melnikov2008} A.\ S.\ Mel'nikov, D.\ A.\ Ryzhov, and M.\ A.\ Silaev, Phys. Rev. B \textbf{78}, 064513 (2008).
\bibitem{Kopnin1995} N.\ B.\ Kopnin and A. V. Lopatin, Phys.\ Rev.\ B \textbf{51}, 15291 (1995).
  
\bibitem{Koulakov1999} A.\ A.\ Koulakov and A.\ I.\ Larkin, Phys.\ Rev.\ B \textbf{60}, 14597 (1999).


\end{thebibliography}
\end{document}